\documentclass[pre,floatfix,twocolumn,showpacs,a4]{revtex4}

\usepackage{graphicx}
\usepackage{dcolumn}
\usepackage{amsmath}
\usepackage{amssymb}
\usepackage{tabularx}
\usepackage{epsfig}
\usepackage{mathrsfs}
\usepackage{rotating}

\begin{document}
\title{Link-space formalism for network analysis}
\author{David M.D. Smith$^{1,2}$}
\email{d.smith3@physics.ox.ac.uk}
\author{Chiu Fan \surname{Lee}$^{1}$}
\author{Jukka-Pekka Onnela$^{1,2,3}$}
\author{Neil F. Johnson$^{4}$}
\affiliation{$^{1}$Physics Department, Clarendon Laboratory, Oxford University, Oxford OX1 3PU, U.K.}
\affiliation{$^{2}$Sa\"id Business School, Oxford University, Oxford OX1 1HP, U.K.}
\affiliation{$^{3}$Laboratory of Computational Engineering, Helsinki University of Technology, Finland}
\affiliation{$^{4}$University of Miami, Coral Gables, Florida FL 33124, USA}

\date{\today}

\begin{abstract}
We introduce the link-space formalism for analyzing network models with degree-degree correlations. The formalism is based on a statistical description of the \emph{fraction of links} $l_{i,j}$ connecting nodes of degrees $i$ and $j$. To demonstrate its use, we apply the framework to some pedagogical network models, namely, random-attachment, Barab\'asi-Albert preferential attachment and the classical Erd\H{o}s and R\'enyi random graph. For these three models the link-space matrix can be solved analytically. We apply the formalism to a simple one-parameter growing network model whose numerical solution exemplifies the effect of degree-degree correlations for the resulting degree distribution. We also employ the formalism to derive the degree distributions of two very simple network decay models, more specifically, that of random link deletion and random node deletion. The formalism allows detailed analysis of the correlations within networks and we also employ it to derive the form of a perfectly non-assortative network for arbitrary degree distribution.
\end{abstract}


\maketitle


Networks -- in particular large networks with many nodes and links -- are attracting widespread attention. The classic reviews \cite{albert:2002,random,Newman:2003} with their primary focus on structural properties have been followed up by more recent ones addressing the role of dynamics, such as spreading and synchronization processes on networks, as well as the role of weights and mesoscopic structures, i.e. cliques and communities, within networks \cite{boccaletti:2006,NewmanBook}. Although several different measures for characterizing networks have been presented, for example in a recent survey \cite{costa:2005}, the simple concept of vertex degree remains unrivalled in its ability to capture fundamental network properties. When comparing the degrees of connected vertices, however, one often finds that they are correlated, a quality that gives rise to a  rich set of phenomena. Degree correlations constitute a central role in network characterization and modelling but, in addition to being important in their own right, also have substantial consequences for dynamical processes unfolding on networks. Given the increasing current interest in network dynamics, understanding structural correlations remains important and timely.

In this paper we provide a detailed mathematical formalism for modelling degree-degree correlations within stochastically evolving, non-equilibrium networks. It is built around a statistical description of inter-node linkages as opposed to single-node degrees. While correlations have been characterized in empirical and model networks, most works devoted to analytical calculations of correlations in models, as pointed out in \cite{barrat:2005}, have been performed only for particular cases. We start by providing a brief overview of degree correlations for network structure and dynamics in Section \ref{sec:Overview} and discuss how the work presented relates to existing studies. The \emph{link-space formalism}, which lies at the core of this paper, is introduced in Section~\ref{sec:NSLS}. The formalism comprises a master equation description of the evolution of a specific matrix construction, termed the \emph{link-space matrix}, which characterizes degree-degree correlations within a network. The formalism can be implemented in a variety of ways. To demonstrate its use, we apply it to two well-known, non-equilibrium examples, namely random-attachment and Barab\'asi-Albert (BA) preferential-attachment networks \cite{Barabasi,Barabasi1} in Section~\ref{sec:NSLS}, and solve the so-called link-space matrix analytically for the steady state of these systems. In Section~\ref{sec:NSLS}, we also apply it to the classical equilibrium random graph of Erd\H{o}s and R\'enyi~\cite{Erdos} (ER) which, interestingly, requires a full, time-dependent solution of the link-space master equations. The cumulative link-space introduced in Section~\ref{sec:NSLS} aids comparison between simulated and analytically-derived link-space matrices and could, in principle, be applied to empirical networks to better ascertain their degree correlations. Analytic derivation of link-space matrices allows detailed analysis of the degree-degree correlations present within networks as is demonstrated in Section~\ref{sec:cor}. Within Section~\ref{sec:cor}, we also show how, for an arbitrary degree distribution, a link-space matrix can be derived which has no correlations present, i.e. the link-space is representative of a perfectly non-assortative network with that degree distribution. We then consider the counter-intuitive prospect of finding steady states of decaying networks in Section \ref{sec:decay}. In Section \ref{sec:mm}, we introduce a simple one-parameter network growth algorithm, which is able to produce networks whose degree distributions exhibit a wide range of power-law exponents via a redirection process similar to the model by Krapivsky and Redner~\cite{Krapivsky}. The model is interesting in its own right in the sense that it makes use of only local information about node degrees. Here the link-space formalism allows us to identify the transition point at which higher exponent degree distributions switch to lower exponent distributions with respect to the BA model. We conclude in Section \ref{sec:conc}. To maintain readability, we postpone the more detailed mathematical derivations to Appendices \ref{app:RA}, \ref{app:BA} and \ref{app:decay}, and refer to them where appropriate.

\section{Overview of degree correlations}
\label{sec:Overview}
A discussion of correlations might naturally start with the phenomenon of \emph{clustering}. The highly influential paper of Barab\'asi ans Albert and many subsequent papers focussed on the modelling and analysis of models which replicate power-law degree distributions observed in many real-world systems \cite{Barabasi, dorogovtsev:2000, Newman:2003}. However, the BA preferential attachment mechanism (discussed in Section~\ref{subsec:BA}) lacks certain key features observed in many of those systems, one such feature being clustering. Clustering is an important local statistic in many networks and reflects the connectedness of neighbours of a node. In a social context, if person $Y$ is friends with both  $X$ and $Z$, one might expect some kind of link between $X$ and $Z$. The measurement of the connectedness of neighbours of nodes is the clustering coefficient. This is often averaged over all nodes and a single value is given as the average clustering coefficient of the network~\cite{random}. Some node labelled $\eta$ in an arbitrary undirected network has $k_\eta$ neighbours, between which there could be a possible $k_\eta(k_\eta-1)/2$ links. If $y$ of these are actually present, the clustering of node $\eta$ is given as $C= 2y/(k_{\eta}(k_{\eta}-1))$. The clustering nature of a network can also be expressed as the average over all nodes of degree $k$ giving a clustering distribution (or \emph{spectrum}), $\overline{C_k}$. As noted by Klemm and Egu\'{i}luz, in a BA network, the average clustering of a given node is independent of its degree \cite{klemm:2002a}, in contrast to the findings of Fronczak {et al.} \cite{Holyst},  and tends to zero in the large-size (thermodynamic) limit. A network model which does exhibit high clustering is that of Watts and Strogatz in which a random rewiring process is carried out on an initially regular lattice~\cite{smallworld}. The networks generated by such a process feature short average shortest path lengths between node pairs (the \emph{small-world} effect). However, these networks do not exhibit a power-law degree distribution. Subsequently, there have been many efforts to build models which can encompass both of these features such as the non-equlibrium growing network model proposed by Klemm and Egu\'{i}luz which features deactivation of a nodes availibility to be connected to \cite{klemm:2002a}. Other models such as that of  Holme and Kim \cite{Holme} and that introduced by  Toivonen \emph{et al.} \cite{toivonen:2006} modified the original BA preferential attachment mechanism, allowing further links between the new node and neighbours of the preferentially selected node. The social network model of Toivonen \emph{et al.} produced communities with dense internal connections.  Szab\'{o} \emph{et al.} formulated a scaling assumption and a mean-field theory of clustering in growing scale-free networks and applied it to the Holme and Kim mechanism \cite{szabo:2003}. As discussed by Bogu\~n\'a and Pastor-Satorras, clustering in networks is closely related to degree correlations \cite{boguna:2003}. Infact, based on the work of Szab\'{o} \emph{et al.}, Barrat and Pastor-Satorras introduced a framework for computing the rate equation for two vertex correlations in the continuous degree and continuous time approximation \cite{barrat:2005}. We shall now describe these degree correlations.

Vertex degree correlations are measures of the statistical dependence of the degrees of neighbouring vertices in a network. In general, $n$-vertex degree correlations, or $n$ point correlations, can be fully characterized by the conditional probability distribution 
$P(k_1,k_2,$ $\ldots,k_n | k=n)$ that a vertex of degree $k=n$ is connected to a set of $n$ vertices with degrees $k_1,k_2, \ldots, k_n$. Two and three-point correlations are of particular interest in complex networks as they can be related to network assortativity and clustering, respectively.  More specifically, two vertex degree correlations (two point correlations) can be expressed as conditional probability $P(k'|k)$ that a vertex of degree $k$ is connected to a vertex of degree $k'$. Similarly, three vertex degree correlations (three point correlations) can be fully characterized by the conditional probability distribution $P(k', k''|k)$ that a vertex of degree $k$ is connected to both a vertex of degree $k'$ and a vertex of degree $k''$. This implies that the degrees of neighbouring nodes are not statistically independent. Reliable estimation of $P(k'|k)$ and $P(k', k''|k)$ requires a large amount of data and, in practice, one often resorts to related measures. Instead of $P(k'|k)$, the average degree of nearest neighbours, $\langle k_{nn}\rangle_k$, of nodes with degree $k$ is often measured. This can be formally related to $P(k'|k)$ \cite{boguna:2003, pastor-satorras:2001}. If $\langle k_{nn}\rangle_k$ increases with $k$, high degree vertices tend to connect to high degree vertices. A network with this property is described as \emph{assortative} or displaying positive degree correlations. If $\langle k_{nn}\rangle_k$ decreases with $k$, high degree vertices tend to connect to low degree vertices (\emph{disassortative} or negatively correlated) \cite{boguna:2003, Newman:2002}. An alternative to $\langle k_{nn}\rangle_k$ is to use a normalised Pearson's correlation coefficient of adjacent vertex degrees providing a single number measure of assortativity as suggested by Newman to further classify networks \cite{Newman:2002, Newman:2003}. These approaches are discussed in more detail in Section~\ref{sec:cor}. To characterize three point correlations, instead of using $P(k', k''|k)$, one can employ the clustering spectrum $\overline{C_k}$, the average clustering coefficient of nodes with degree $k$,  which can be related to $P(k',k''|k)$ \cite{boguna:2003}. In many real-world networks such as the Internet~\cite{pastor-satorras:2001}, the clustering spectrum is a decreasing function of degree and while this is sometimes interpreted as a signature of hierarchical structure in a network, Soffer and V\'azquez suggested that this is a consequence of degree-degree (two point) correlations that enter the definition of the standard clustering coefficient \cite{soffer:2005}. The authors introduced a different definition for the clustering coefficient that does not have the degree-correlation `bias', i.e. a three point correlation measure that filters out two point correlations. Following the suggestion of Maslov {et al.} \cite{maslov:internet} that these phenomena might arise from topological constraints rather than evolutionary mechanisms, Park and Newman demonstrated that dissasortative degree correlations observed in the Internet could be explained via the restriction of there being no double edges between nodes \cite{park:2003}. In contrast, social networks have been found to be assortative \cite{toivonen:2006}. Similarly, Catazaro \emph{et al.} observed that the network of scientific collaborations was assortative and presented a model to reproduce this feature \cite{caldarelli:2004}. 

Functional processes occurring on networks are influenced by degree correlations, highlighting the importance of their role in complex networks. 
Egu\'iluz and Klemm considered highly clustered scale-free networks and showed that correlations play an important role in epidemic spreading \cite{equiluz:2002}. The time average of the fraction of infected individuals in the steady-state undergoes a phase transition at a finite critical infection probability. They related this critical threshold to the transmission probability and the mean degree of nearest neighbours of all nodes in the system, $\langle k_{nn}\rangle$, the conjectured criterion for epidemic spreading being related to the product of the two. The value $\langle k_{nn}\rangle$ scales with the system size in their highly clustered scale-free network more slowly than in the random scale-free which is a byproduct of the dissasortativity of the system. Consequently, whereas in random scale-free networks in which viruses with extremely low spreading probabilities can prevail, the absence of connections between highly connected nodes in highly clustered scale-free networks protects the system against epidemics \cite{equiluz:2002}. Interestingly, Bog\~un\'a \emph{et al.} assert that any scale-free network of appropriate exponent will have diverging $\langle k_{nn}\rangle$ in the thermodynamic limit, resulting in no threshold properties for epidemic spreading regardless of the correlations within the network \cite{boguna:absence} in contrast to the earlier suggestion of Bog\~un\'a and Pastor-Satorras \cite{boguna:2002}. Brede and Sinha induced correlations to Erd\H{o}s-R\'enyi and scale-free networks by rewiring them appropriately to examine their dynamic stability. They mapped the adjacency matrices into Jacobian matrices and examined the largest eigenvectors, reflecting the decay rates of perturbations about an equilibrium state. They found that positive correlations reduced their dynamical stability \cite{brede:2005}. Similarly, Bernardo \emph{et al.} induced negative degree correlations in scale-free networks whose links couple non-linear oscillators. Through analysis of the eigenratio of the Laplacians of such networks, they found that network synchronisability improved as the network was made more dissassortative \cite{bernardo:2005a}. Based on this result, they conjectured that negative degree correlations may emerge spontaneously as the networked system attempts to become more stable \cite{bernardo:2005}. The same authors found similar results to hold also for weighted networks \cite{bernardo:2005a}. Maslov and Sneppen found that in ``interaction and regulatory networks, links between highly connected proteins were systematically suppressed, whereas those between highly connected and low-connected pairs of proteins were favoured", a topological organisation that increases the overall robustness of the network to perturbations through ``localising deletous perturbations"  \cite{maslov:2002}. This is consistent with the findings of Berg \emph{et al.} whose model of the evolution of protein interaction networks exhibited disassortativity consistent with their empirical findings \cite{berg:2004}.

Krapivsky \emph{et al.} used a master equation method, in which rate equations for the densities of nodes of a given degree are employed, to investigate the steady state of the BA preferential attachment mechanism~\cite{KrapivskyDeriv}. As illustrated in Section~\ref{sec:NSLS}, this is a general method which can be applied to various growing network models. To extend the method to encompass two-point correlations, Krapivsky and Redner applied the approach to the number $N_{k,l}$ of nodes with total degree $k$ connected to ancestor nodes of total degree $l$ in a \emph{directed} network~\cite{Krapivsky}. They solved analytically the master equations for the steady-state of a specific directed network growth model, namely the Growing with Redirection algorithm~\cite{Krapivsky}, similar to the mixture model introduced in Section~\ref{sec:mm}. Bogu\~n\'a and Pastor-Satorras considered a link orientated description of two point correlations. For \emph{undirected} graphs, they introduced a symmetric matrix whose elements $E_{k,l}$ represent the number of links connecting nodes degree $k$ to nodes degree $l$~\cite{boguna:2003}. Bogu\~n\'a and Pastor-Satorras related this matrix to the joint degree distribution $P(k'\mid k)$ but noted that finite size effects make empirical evaluation of the matrix difficult, suggesting the use of the spectrum $\langle k_{nn} \rangle_k$ as a more suitable observable. Here, we introduce a similar matrix construction which can be applied to both the undirected \emph{and} directed scenarios. We call this the \emph{link-space} matrix. We show that it is possible to construct master equations to model the evolution of this matrix which can be applied to a wide variety of network evolution algorithms retaining important degree correlations which can be critical to the network's development. This framework is termed the \emph{link-space formalism} and is introduced in Section~\ref{sec:NSLS}. In certain cases, these master equations can be solved analytically providing a full time-dependent solution or a steady-state solution of the link-space matrix of the network. From these analytically derived link-space matrices both the degree and joint degree distributions can be obtained allowing accurate analysis of degree correlations. The formalism also allows the derivation of the form of networks of predetermined correlation properties such as a perfectly non-assortative network (Section~\ref{subsection:nonassort}) and the derivation of the steady-state solutions to network decay algorithms (Section~\ref{sec:decay}). The formalism can also be employed in its iterative guise to provide approximations to the steady-state when an analytic solution is not possible such as for the mixture model of Section~\ref{sec:mm}.

One and two-point correlations have natural physical counterparts within the network, specifically the nodes and links. The one-point correlations $P(k)$ are simply related to the fraction of nodes in the network with degree $k$. As described in detail within this paper, two-point correlations $P(k'\mid k)$ are related to the fraction of links within a network connecting nodes of degree $k$ with nodes of degree $k'$, hence the term \emph{link-space}. One can extend the analysis to three-point correlations, $P(k',k''\mid k)$. This would be related to the number (fraction) of pairs of links within a network sharing a common node of degree $k$, the remaining link ends being connected to nodes of degrees $k'$ and $k''$ and all possible `open triangles' within the network would have to be considered. Clearly, the process can be extended to arbitrary $n$ point correlations although the physical interpretation of the appropriate measurable quantities will become increasingly obscure.

\section{Node-space and link-space}
\label{sec:NSLS}
We now introduce the link-space formalism. Consider a simple, growing, non-equilibrium network in which one node is added to a network at each timestep and this node is connected to the existing network with exactly $m$ undirected links. The process is governed by an attachment probability \emph{kernel} $\Theta_j$, defined as the probability that a specific, newly-introduced link attaches to any node of degree $j$ within the existing network. At some time $t$ there exist $X_i(t)$ nodes of degree $i$ and we wish to compute the expected number of nodes with degree $i$ at time $t+1$. The \emph{node-space master equations} can be expressed in terms of the attachment kernels and are written
\begin{eqnarray}
\langle X_i(t+1) \rangle  &=& X_i(t) +m\Theta_{i-1}(t)-m\Theta_{i}(t) \, \, \,i>m,\nonumber\\
\langle X_m(t+1)\rangle &=& X_m(t) + 1 - m\Theta_m(t),
\label{eqn:nodeXmaster}
\end{eqnarray}
since a new node of degree $m$ is added to the existing network at each timestep and there are no nodes with degree less than $m$. 

So far we have said nothing about the attachment mechanism, and have made
the easily geneneralizable restriction that only one node is being added per timestep with undirected links. We now follow a similar analysis, but retain the node-node linkage 
correlations that are inherent in many real-world systems \cite{Newman:2002, Callaway,Krapivsky}.
Consider any link in a general network -- we can describe it by the degrees of the two nodes that it connects. 
Hence we can construct a matrix $\mathbf{L}(t)$ such that the element $L_{i,j}(t)$ is equal to the number of links from nodes of degree $i$ to nodes of degree $j$ for $i\ne j$ at some time $t$. To ease the mathematical analysis below, the diagonal element $L_{i,i}(t)$ is defined to be \emph{twice} the number of links between nodes of degree $i$ for the undirected graph, a mathematical convienience also observed by Bogu\~n\'a and Pastor-Satorras~\cite{boguna:2003}. For undirected networks $\mathbf{L}(t)$ is symmetric and $\sum_{i,j} L_{i,j}(t) = 2M(t)$, twice the total number of links $M(t)$ in the network which is simply $mt$ when introducing $m$ links per timestep. The matrix, $\mathbf{L}$, represents a surface describing degree-degree correlations in the network (see Figs.~\ref{fig:RA}, \ref{fig:BA} and \ref{fig:ER}) and is called the \emph{link-space} matrix.

Consider one of the newly introduced links, one end of which is attached to the new node. The probability of selecting any node of degree $i-1$ within the existing network for the other end to attach to is given by the attachment probability $\Theta_{i-1}(t)$. Suppose an $i-1$ node is selected. The fraction of nodes of degree $i-1$ that are connected to nodes of degree $j$ is 
\begin{displaymath}
\frac{L_{i-1,j}(t)}{(i-1)X_{i-1}(t)}.
\end{displaymath}
The expected increase in links from nodes of degree $i$ to nodes of degree $j$, through the 
attachment of the new node to a node of degree $i-1$, is given by 
\begin{displaymath}
\frac{\Theta_{i-1}(t) L_{i-1,j}(t)} { X_{i-1}(t)}.
\end{displaymath}
Since each link has two ends, the value $L_{i,j}$ can increase by a connection to an $(i-1)$-degree node which is in turn connected to a $j$-degree node, or by connection to an $(j-1)$-degree node which is in turn connected to an $i$-degree node. We write master equations governing the evolution of the link-space matrix as the evolution of the expected number of links from $i$ to $j$ degree nodes, i.e. the number of $i \leftrightarrow j$ links. This is the \emph{link-space formalism} and is written for the (generalizable) case of adding one new node with $m$ undirected links to the existing network as
 \begin{eqnarray}
\langle L_{i,j}(t+1)\rangle  & =& L_{i,j}(t)  
+\frac{m\Theta_{i-1}(t)L_{i-1,j}(t)}{X_{i-1}(t)}\nonumber \\
{}&{}&+\frac{m\Theta_{j-1}(t)L_{i,j-1}(t)}{X_{j-1}(t)} -~\frac{m\Theta_{i}(t)L_{i,j}(t)}{X_{i}(t)}\nonumber \\
{}&{}& - \frac{m\Theta_{j}(t)L_{i,j}(t)}{X_{j}(t)}, \, \, \, i,j>m, \nonumber \\
\langle L_{m,j}(t+1)\rangle & =& L_{m,j}(t) + m\Theta_{j-1}(t)  +\frac{m\Theta_{j-1}(t)L_{m,j-1}(t)}{X_{j-1}(t)} \nonumber\\
{}&{} &- \frac{m\Theta_{m}(t)L_{m,j}(t)}{ X_{m}(t)}\nonumber \\
{}&{}&-\frac{m\Theta_{j}(t)L_{m,j}(t)}{ X_{j}(t)}, \, \, \,  j>m.
\label{eqn:lspace}
\end{eqnarray}

There are a variety of ways in which both the node-space and link-space master equations can be approached. For example, a full, time-dependent solution could be investigated as in Section~\ref{subsec:ER} or, using appropriate initial conditions, the equations can be iterated over the required timescale as in Section~\ref{sec:mm}. We can also investigate the possibility of a steady state of the algorithm under scrutiny. To do so, we assume that there exists a steady state in which the degree distribution remains static and investigate a solution~\footnote{The master equation does not represent the evolution of an ensemble average of networks. Each specific realisation will have its own evolution of attachment kernel which cannot be described by the ensemble average. The existence of a steady state to the master equation does not necessarily imply that specific realisations converge upon it in the large $N$ limit but simply that the solution is static with respect to the growth algorithm. In practice, however, a master equation approach often yields good analytic agreement with even single realisations~\cite{David}.}. Under this assumption, the fraction of nodes $c_i(t) = X_i(t)/N(t)$ which have a given degree remains constant such that $\langle X_i(t+1) \rangle - X_i(t) \approx {dX_i}/{dt} = c_i $ when one new node is added per timestep ($N(t)=t$). It is also assumed that in the steady state, the attachment kernels are static too. We drop the notation `$(t)$' to indicate the steady-state and can rewrite Eq.~(\ref{eqn:nodeXmaster}) as
\begin{eqnarray}
c_i &=&m\Theta_{i-1}-m\Theta_{i}, \,\,\, i>m,\nonumber\\
c_m &=&1-m\Theta_m.
\label{eqn:nsmaster}
\end{eqnarray}
The fraction of links between nodes of degree $i$ and nodes of degree $j$ can be expressed as the normalized link-space matrix, $l_{i,j}(t) = L_{i,j}(t)/M(t)$, which sums to $2$ in the undirected case. In the steady state we can assume that these values are static and can rewrite the link-space master equation (Eq.~(\ref{eqn:lspace})) as 
\begin{eqnarray}
l_{i,j}& =& \frac{\frac{\Theta_{i-1}}{c_{i-1}}l_{i-1,j} ~+~ 
\frac{\Theta_{j-1}}{c_{j-1}}l_{i,j-1}}{\frac{1}{m}~+ 
\frac{\Theta_{i}}{c_{i}}~+~\frac{\Theta_{j}}{c_{j}}  },\, \, \,i,j>m,  
\nonumber\\
l_{m,j}& =& \frac{\frac{\Theta_{j-1}}{c_{j-1}}l_{m,j-1}~+~\frac{\Theta_{j-1}}{m}}{\frac{1}{m}~+ 
\frac{\Theta_{m}}{c_m}~+~\frac{\Theta_{j}}{c_{j}} }, \, \, \,j>m.
\label{eqn:linkspacemaster}
\end{eqnarray}
The notation `$(t)$' has again been dropped to indicate the steady-state and here the generalizable situation of adding one new node per timestep is considered.

To apply the link-space formalism one starts by specifying the model dependent attachment kernel $\Theta_i$. To investigate a time-dependent solution, the attachment kernel is substituted into Eqs.~(\ref{eqn:nodeXmaster}) and (\ref{eqn:lspace}). To investigate a steady-state solution, the kernel is substituted into the Eqs.~(\ref{eqn:nsmaster}) and (\ref{eqn:linkspacemaster}) to yield recurrence relations for $c_i$ and $l_{i,j}$ respectively which can be solved analytically in some cases. The number of $i$-degree nodes is given by $X_i(t) = i^{-1} \sum_k L_{i,k}(t)$ which allows us to retrieve the degree distribution from the normalized link-space matrix:
\begin{eqnarray}\label{eqn:dist}
c_i(t) &=& \frac{M(t) \sum_{k=1}^{\infty} l_{i,k}(t)}{i N(t)}. 
\end{eqnarray}

Degree distributions of empirically observed or simulated networks typically become dominated by noise at large degrees reflecting the small probabilities associated with these values occurring. In the link-space, the situation is exacerbated and the high $i,j$ limit reflects connections between these high degree nodes. This is, of course, rarer than the existence of nodes of either degree. Following the conventional approach which is applied to degree distributions~\cite{random}, we can use a cumulative representation of the link-space to address this issue. The use of a cumulative binning technique averages over stochasticity in the system. With degree distributions, regression techniques (curve fitting) applied to the cumulative distribution is used to obtain a more accurate description of the actual degree distribution than would be obtained from fitting to the empirical distribution itself~\cite{random}. The process can be similarly applied in the link-space. Surface fitting could be applied to the cumulative link-space obtained from an empirical network. From the cumulative fitted surface, a more accurate representation of the actual link-space could be obtained for the empirical network, from which a better representation of the network's correlations could be obtained. This process also allows for comparison between simulated and analytically-derived link-space matrices. We define the cumulative link-space matrix, $_{cum}l_{i,j}$ to be
\begin{eqnarray}
_{cum}l_{i,j}&=&\sum_{x=i}^\infty \sum_{y=j}^\infty l_{x,y}.
\end{eqnarray}
Note that we have not lost generality in that given a cumulative linkspace matrix, the actual link-space can be derived using variations on the following:
\begin{eqnarray}
l_{i,j}&=&_{cum}l_{i,j} - _{cum}l_{i+1,j}-_{cum}l_{i,j+1}+_{cum}l_{i+1,j+1}.
\end{eqnarray}
The computation involved when evaluating $_{cum}l_{i,j}$ can be cut down considerably by first evaluating $_{cum}l_{1,1}$, which in the undirected graph is equal to $2$. The leftmost column (or top row) for $i>1$ can then be evaluated as
\begin{eqnarray}
_{cum}l_{i,1}&=& _{cum}l_{i-1,1}-\sum_{x=1}^\infty l_{i-1,x}.
\end{eqnarray}
Note that the second term on the right hand side is a row sum of the normalized link-space matrix and, hence, quickly evaluated. Indeed this row sum can be  related to the degree distribution $c_{i-1}$ from Eq.~(\ref{eqn:dist}). Subsequent elements can be evaluated using the following simple identity,
\begin{eqnarray}
_{cum}l_{i,j}&=&_{cum}l_{i-1,j}+_{cum}l_{i,j-1}\nonumber \\
{}&{}& - _{cum}l_{i-1,j-1}+l_{i-1,j-1}.
\end{eqnarray}

We will now demonstrate the use of the link-space formalism to study a random attachment model and the Barab\'asi-Albert (BA) model using steady-state solutions and the classical Erd\H{o}s and R\'enyi random graph using a time-dependent solution.

\subsection{Random attachment model (steady-state solution).}
\label{subsec:RA} 
\begin{figure*}
\begin{center}
\includegraphics[width=0.45\textwidth]{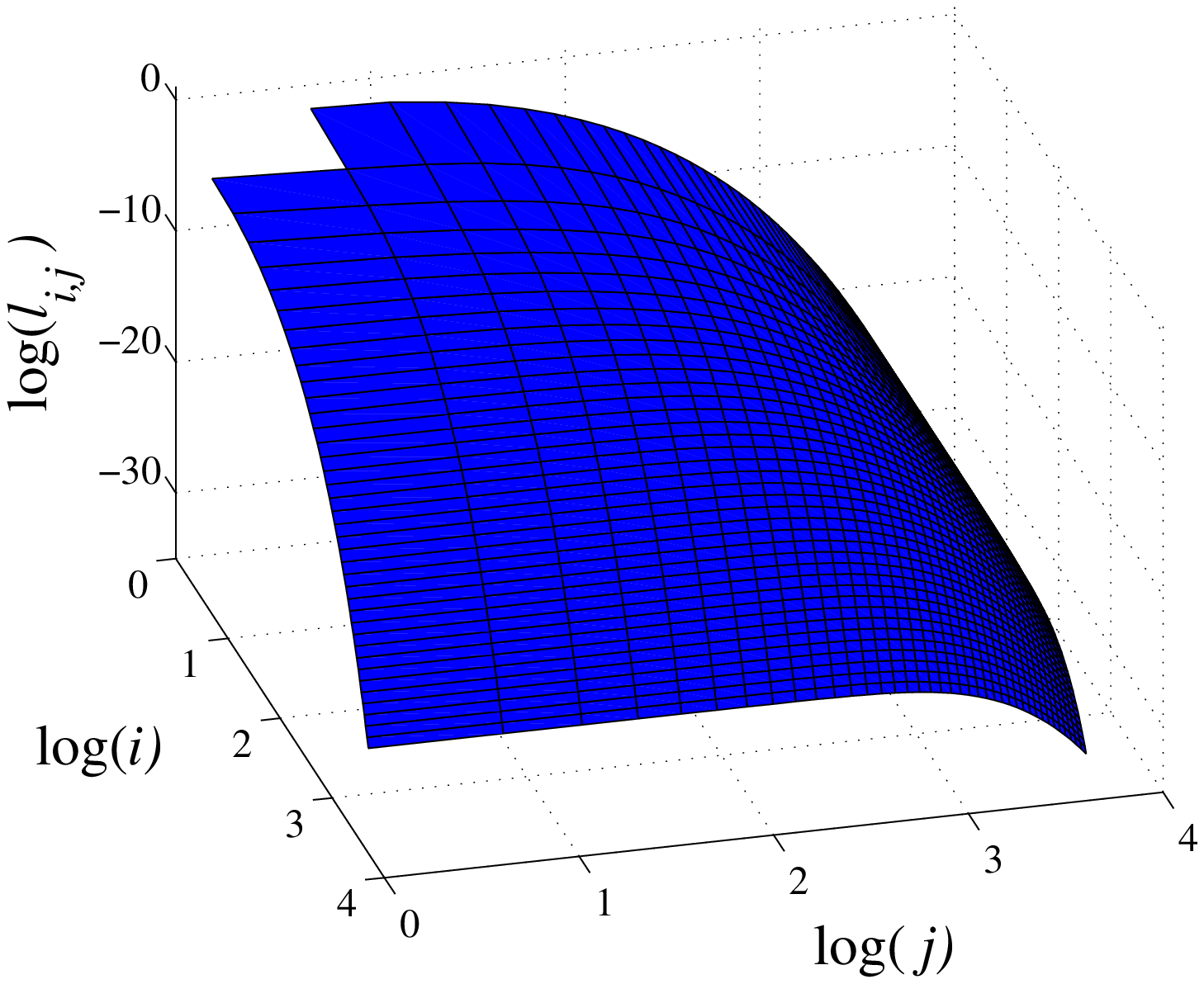}
\includegraphics[width=0.45\textwidth]{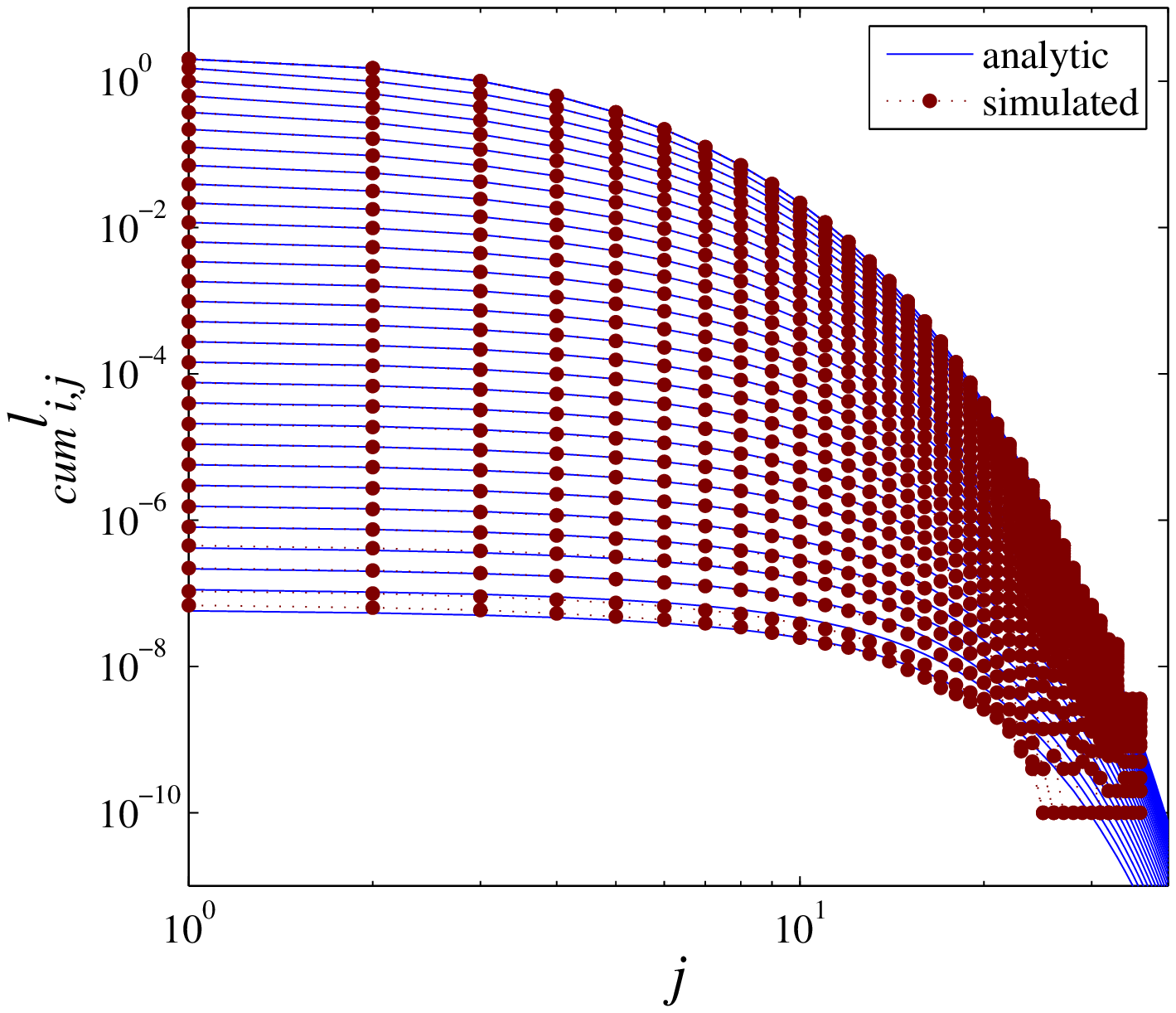}
\caption{ \label{fig:RA} (Color Online) Left: the analytically-derived, normalized link-space matrix for the random attachment growth algorithm. This could be filled to arbitrary size but is truncated to maximum degree of $40$ here. Right: comparison of the cumulative link-space matrices for the analytic solution and a simulation of the algorithm. The simulation comprises an ensemble average of $100$ networks grown to $10^8$ nodes. The maximum degree obtained was $37$. The first $30$ rows ($i$ values) of the cumulative link-space are illustrated and finite-size effects are noticable at high $j$. }
\end{center} 
\end{figure*}

Consider first a random-attachment model in which at each timestep  a new node is added to the network and connected to an existing node with uniform probability without any preference (`Model A' in \cite{Barabasi,Barabasi1}) with one undirected link such that $m=1$ and $M(t)\approx N(t)= t$. We assume a steady-state solution and the attachment kernel is $\Theta_i= X_i/t =c_i$. Substituting into Eq.~(\ref{eqn:nsmaster}), we obtain the recurrence relation $c_{i+1} = c_i / 2$, which yields the familiar degree distribution $c_i = 2^{-i}$. Substituting into Eq.~(\ref{eqn:linkspacemaster}) yields the recurrence relation
\begin{eqnarray}\label{eqn:ramaster2}
l_{i,j}&=& (l_{i-1,j}+l_{i,j-1}) / 3, \, \, \, \, i,j~>1,\nonumber\\
l_{1,j}&=& (c_{j-1}+l_{1,j-1}) / 3,\, \, \, \, j>1,
\end{eqnarray}
with $l_{1,1}=0$. The exact solution for $l_{i,j}$ is
\begin{eqnarray}
l_{i,j} &= &\sum_{x=2}^{j} 
\frac{ { i-1+j-x \choose j-x }}{3^{(i+j-x)}2^{(x-1)}} 
 + \sum_{x=2}^{i} 
\frac{   {(i-1+j-x) \choose i-x} }{3^{(i+j-x)}2^{(x-1)}}, \, \, \, \, i,j~>1,\nonumber\\
l_{1,j} &= &\sum_{k=1}^{j-1} (3^k 2^{j-k})^{-1},\, \, \, \, j>1,
\end{eqnarray}
where ${x \choose y}$ is the conventional combinatorial `choose' function. Further mathematical details are given in Appendix \ref{app:RA}. We shall make use of this solution to investigate the correlations of such a network in Section \ref{sec:cor}. This normalized link-space matrix is illustrated in Fig.~\ref{fig:RA} along with a comparison to simulated networks using the cumulative link-space matrix. 

\subsection{Barab\'asi-Albert (BA) model (steady-state solution).}
\label{subsec:BA}
\begin{figure*}
\begin{center}
\includegraphics[width=0.45\textwidth]{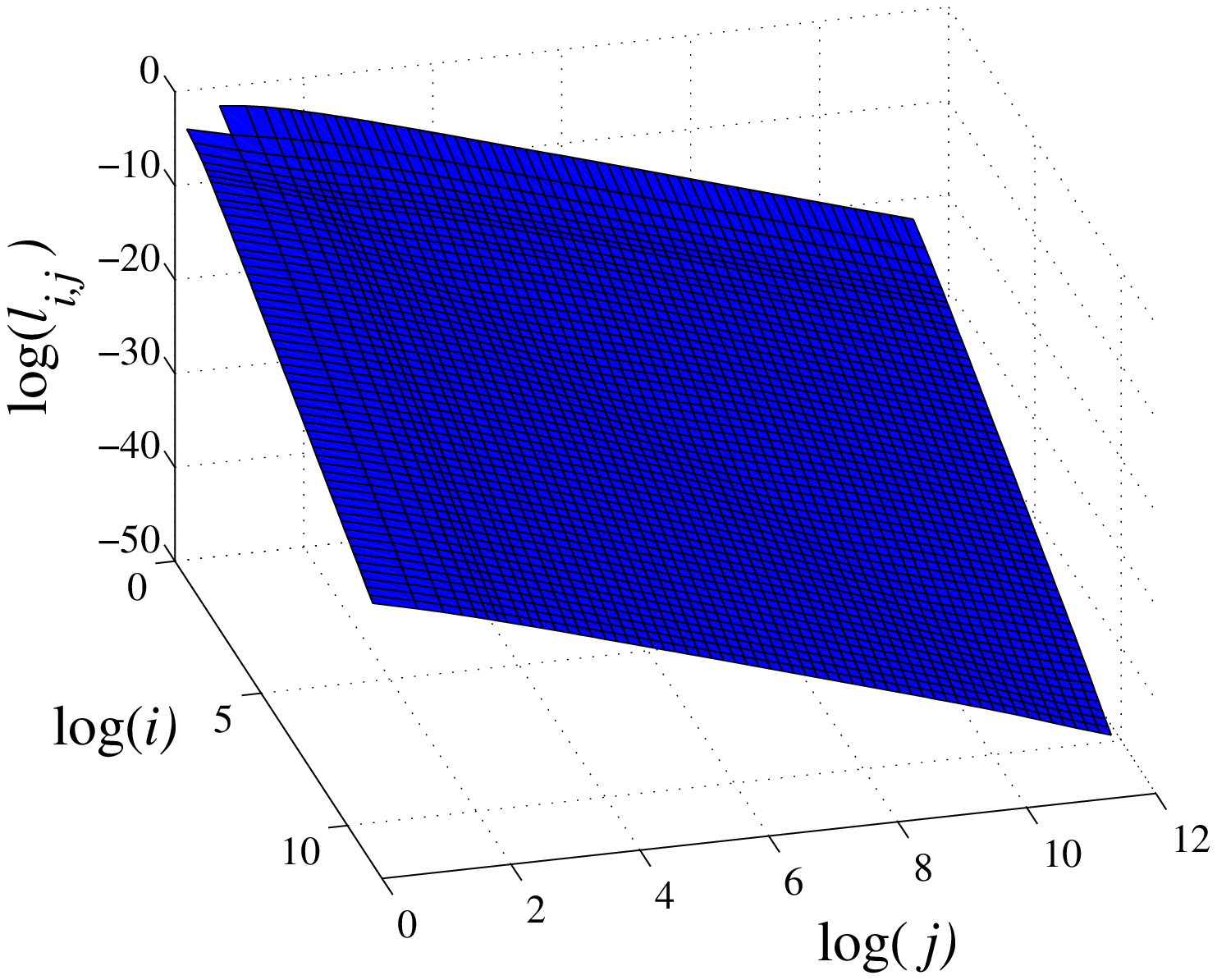}
\includegraphics[width=0.45\textwidth]{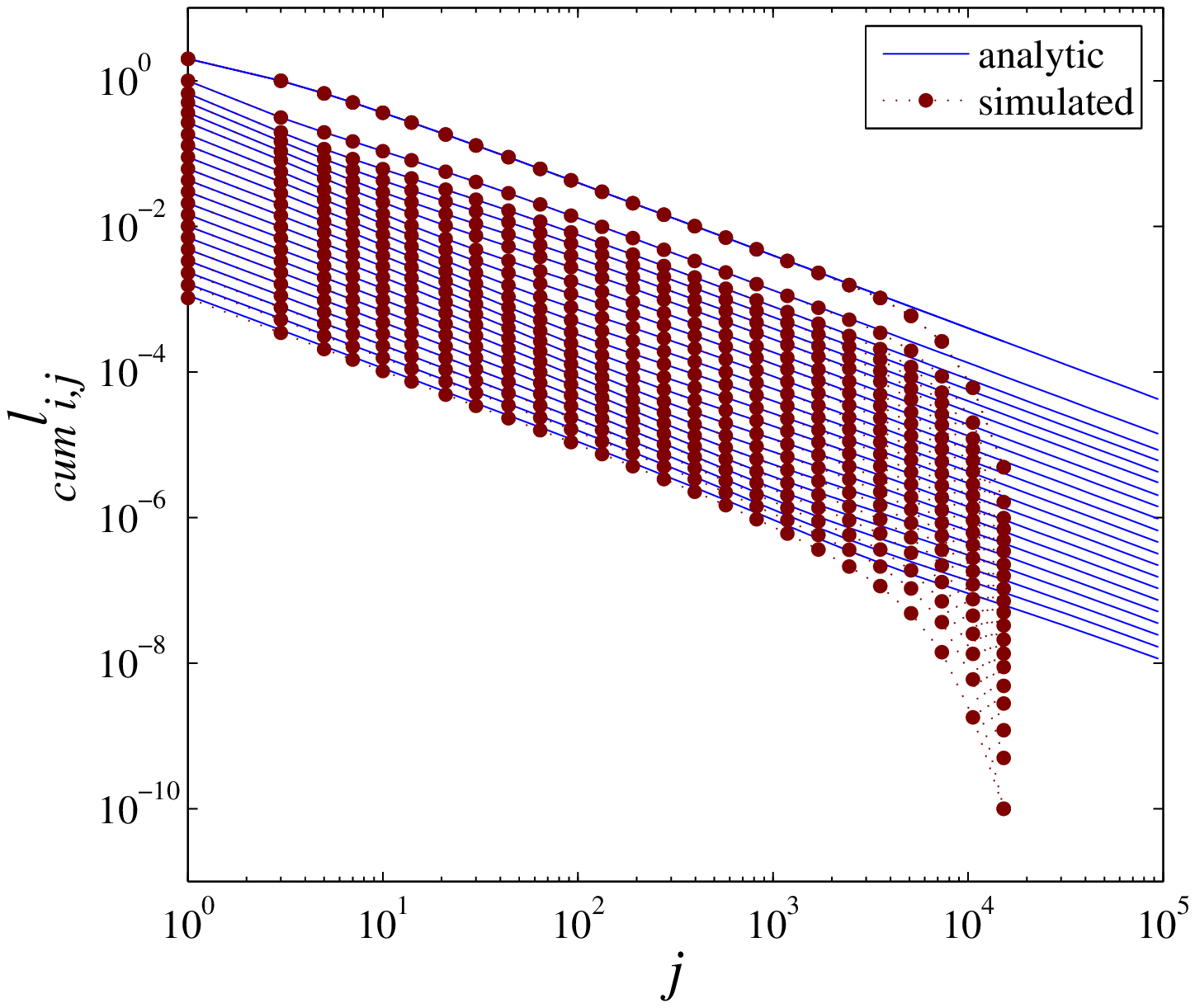}
\caption{ \label{fig:BA} (Color Online) Left: the analytically-derived, normalized link-space matrix for the preferential attachment growth algorithm. This could be filled to arbitrary size but is here truncated to maximum degree of approximately $10^5$. Right: comparison of the cumulative link-space matrices for the analytic solution and a simulation of the algorithm. The simulation comprises an ensemble average of $1000$ networks each grown to $10^7$ nodes. The maximum degree obtained was $17609$. Various rows ($i$ values) from $1$ to $3500$ of the cumulative link-space are illustrated and the effects of the finite nature of the simulation are apparent. }
\end{center} 
\end{figure*}

In the BA model \cite{Barabasi,Barabasi1}, at each timestep a new node is added to the network and connected to $m$ existing nodes with probabilities  proportional to the degrees of those nodes, i.e. $\Theta_i \propto i$ yielding 
\begin{equation}
\label{eqn:SF}
\Theta_i(t) = \frac{i X_i(t)}{\sum_{j} j X_j(t)} = \frac{i X_i(t)}{2M(t)} \approx \frac{i X_i(t)}{2mt}. 
\end{equation}
We consider the scenario when the new node is added with one undirected link, $m=1$, and the attachment kernel is well approximated by 
$\Theta_i \approx  i c_i / 2$. 
Substituting this into Eq.~(\ref{eqn:nsmaster}) yields the recurrence relation 
$c_{i}=\frac{(i-1) c_{i-1}}{2}-\frac{i c_i}{2}
=\frac{i-1}{i+2}c_{i-1}$,
whose solution is $c_{i}=\frac{4}{i(i+1)(i+2)}$.
Using the same substitution, the link-space master equations yield the recurrence relations:
\begin{eqnarray}\label{eqn:sfmaster3}
l_{i,j}&=& \frac{(i-1)l_{i-1,j}~+~(j-1)l_{i,j-1}}{2+i+j},\ \ \ i,j>1,\nonumber\\
l_{1,j}&=&\frac{(j-1)c_{j-1}~+~(j-1)l_{1,j-1}}{3+j},\ \ \ j>1.
\end{eqnarray}
The exact solution for $l_{i,j}$ is obtained by algebraic manipulation of 
the normalised link-space matrix 
using the previously derived degree distribution (further mathematical details are given in Appendix \ref{app:BA}) and is given by
\begin{eqnarray}
l_{i,j}&=& \frac{4(j-1)!(i-1)!}{(j+i+2)!}\Big\{G(i+1)+2G(i)-3G(i-1) \nonumber\\ 
{}&{}& +\frac{1}{2}\sum_{x=2}^{i}(x-1)(x+6)\left[G(i-x)-G(i-x-1)\right] \Big\},\nonumber\\
{}&{}&{} 
\end{eqnarray}
where $G(x)= \left\{\begin{array}{l r}
\frac{(j+x-1)!}{x!(j-1)!} & \textrm{for $x\ge0$}\\
0 & \textrm{~~~~~~~~for $x<0$}. \end{array}\right. $\\
The first few rows of this matrix have the form:
\begin{eqnarray}
l_{1,j}& =& \frac{2(j+6)(j-1)}{j(j+1)(j+2)(j+3)}, \nonumber \\
l_{2,j}&=&\frac{2j(j-1)(j+10)+48}{3j(j+1)(j+2)(j+3)(j+4)}, \nonumber\\
l_{3,j}&=&\frac{3j^4+42j^3-3j^2+246j+360}{9j(j+1)(j+2)(j+3)(j+4)(j+5)}.
\label{eqn:sfexact}
\end{eqnarray}
We shall use this solution to investigate the correlations with this network in Section \ref{sec:cor}. This normalized link-space matrix is illustrated in Fig.~\ref{fig:BA} along with a comparison to simulated networks. 

\subsection{Erd\H{o}s and R\'enyi random graph (time-dependent solution).}
\label{subsec:ER}
The Erd\H{o}s and R\'enyi (ER) classical random graph is the quintessential equilibrium network model~\cite{Erdos, random}. However, to employ the link-space formalism, which tracks the evolution of a network's correlation properties, we must model it as an evolving, non-equilibrium network. To do so is a straightforward process and the model procedes as follows. At each timestep, we add one new node to the existing network. All possible links between the new node and \emph{all} existing nodes are considered and each is established with probability $\alpha$. That is, a biased coin toss (Bernoulli trial) is employed for every node within the existing network to decide whether a link is formed between it and the new node. Conseqently the expected number of new, undirected links with which the new node connects to the existing network is $\langle m(t) \rangle  = \alpha N(t-1) \approx \alpha t$ and this would be described as an accelerating network~\cite{acc}. No new links are formed between existing nodes. This model subsequently produces a network in which the probability of a link existing between any pair of nodes is simply $\alpha$ and is thus representative of the ER random graph~\cite{Erdos}. A network grown to $t$ nodes will have mean degree $ \alpha t$ which will also be the expected degree of all nodes in the network. Whilst the random attachment and preferential attachment models both have steady-state assymptotic behaviour, clearly this model doesn't and a full, time-dependent solution is required.

\begin{figure*}
\begin{center}
\includegraphics[width=0.45\textwidth]{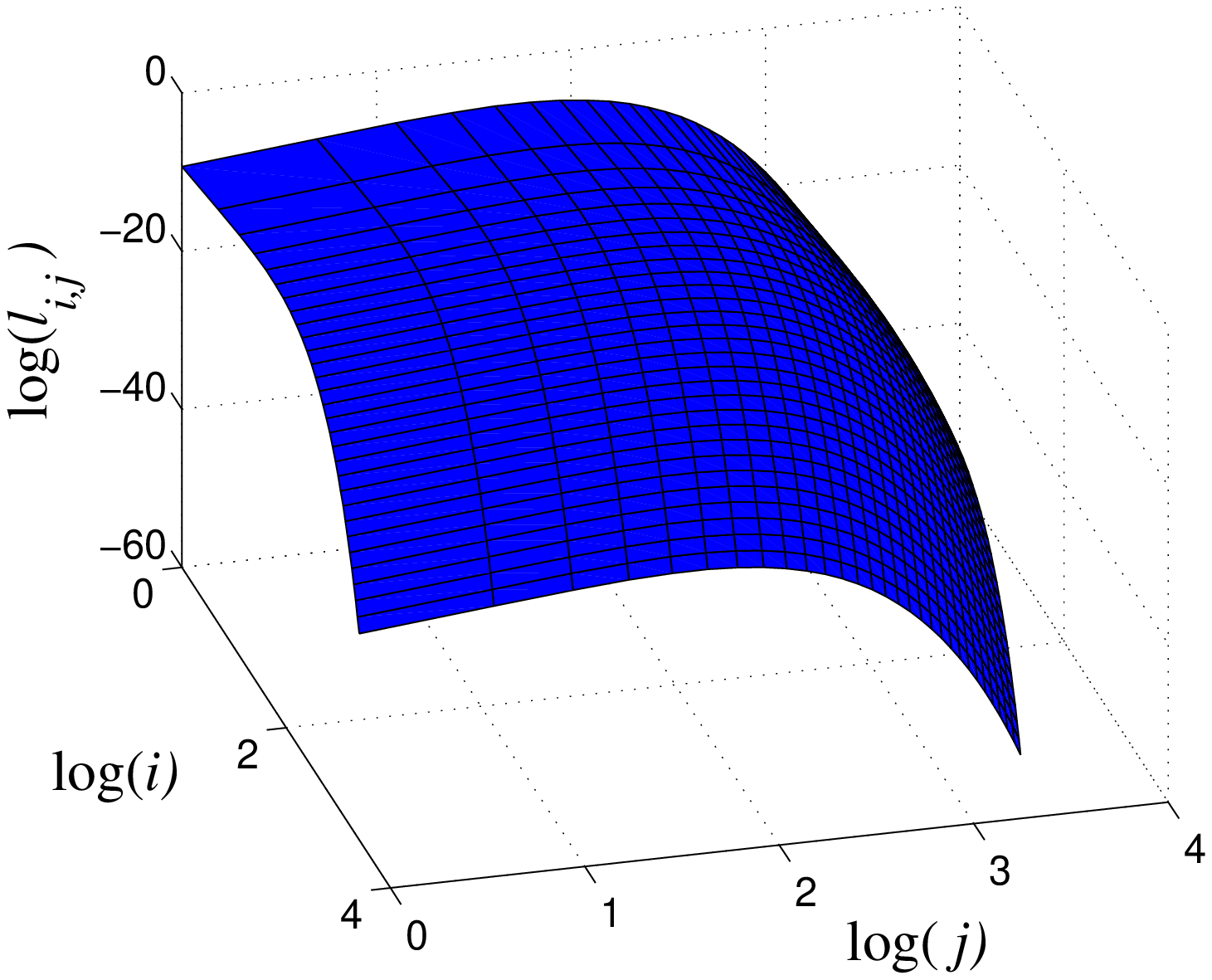}
\includegraphics[width=0.45\textwidth]{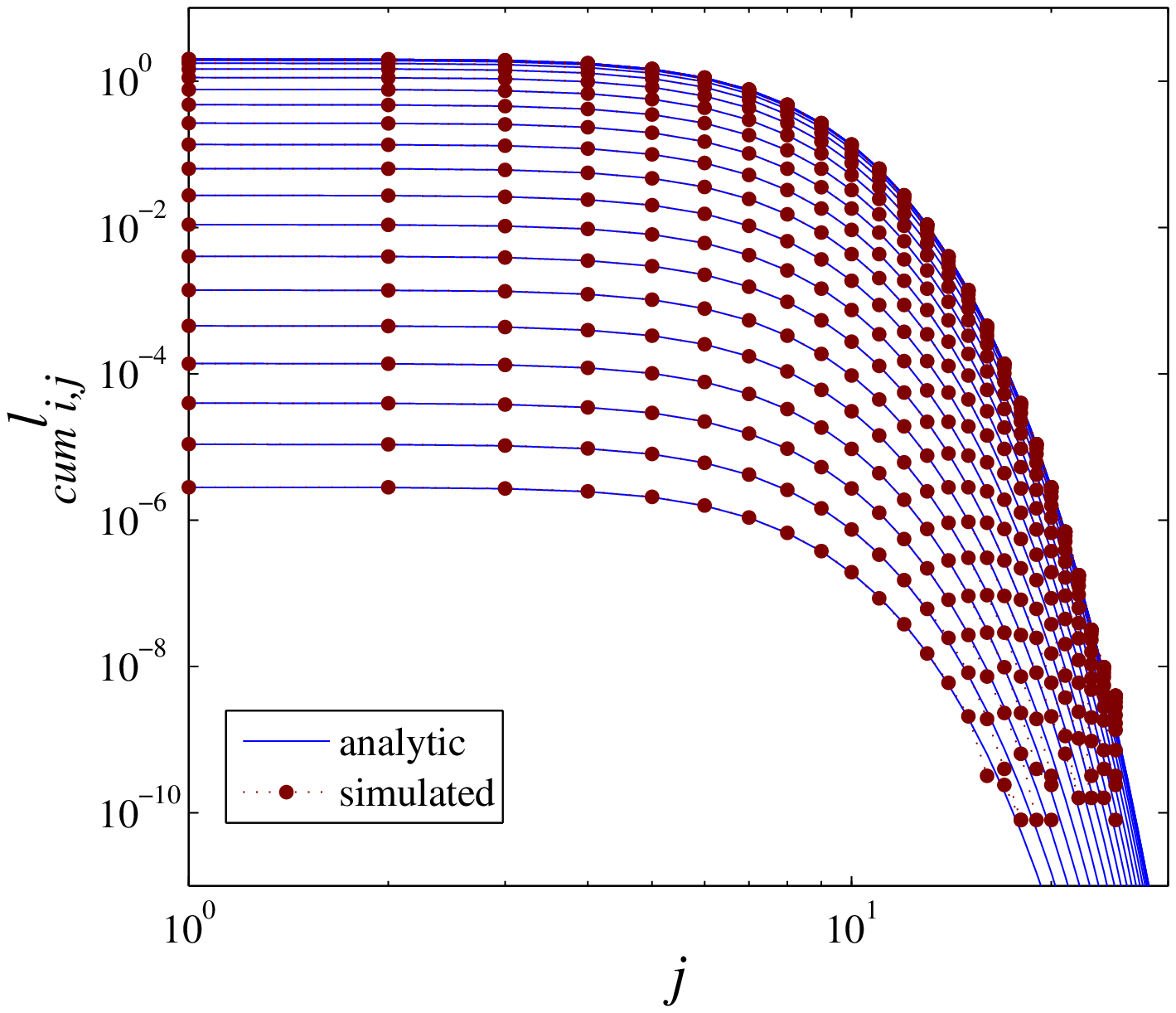}
\caption{ \label{fig:ER} (Color Online) Left: the analytically-derived, normalized link-space matrix for the Erd\H{o}s and R\'enyi (ER) random graph with mean degree equal to $5$. This could be filled to arbitrary size but is here truncated to maximum degree of $30$. Right: comparison of the cumulative link-space matrices for the analytic solution and a simulation of the algorithm. The simulation comprises an ensemble average of $500$ networks each consisting of $10^7$ nodes. The probability that a link exists between any pair of nodes is $\alpha = 5\times10^{-7}$ and the maximum degree obtained was $25$. The first $20$ rows ($i$ values) of the cumulative link-space are illustrated and, again, the finite nature of the simulation is apparent for high $j$.}
\end{center} 
\end{figure*}

Using the random attachment probability kernel $\Theta_i(t) = X_i(t)/N(t)$, the node-space master equation for the number of nodes of degree $i$ can be written
\begin{eqnarray}\label{eqn:acc1}
\langle X_i(t+1)\rangle&=&X_i(t)+\frac{\langle m(t) \rangle X_{i-1}(t)}{N(t)} -\frac{\langle m(t) \rangle X_{i}(t)}{N(t)}\nonumber \\
{}&{}& +P\{m(t)=i\}.
\end{eqnarray}
The second term on the right hand side reflects the expected number of connections to $i-1$ degree nodes making them nodes of degree $i$. The last term is the probability that the new node itself is a node of degree $i$. This will be binomially distributed and, if we assume that $t$ is large, we can make a Poisson approximation. Recalling that the fraction of nodes of degree $i$ is given by $c_i(t) = X_i(t)/N(t) = X_i(t)/t$, we can rewrite Eq.~(\ref{eqn:acc1}) as
\begin{eqnarray}\label{eqn:acc2}
\frac{\textrm{d} \left( tc_i(t) \right)}{\textrm{d} t}&=& \alpha c_{i-1}(t) -\alpha c_i(t) + \frac{e^{-\alpha t}(\alpha t)^i}{i!}.
\end{eqnarray}
The time-dependent solution of Eq.~(\ref{eqn:acc2}) yields the degree distribution for this model and is simply 
\begin{eqnarray}
c_i(t)&=& \frac{e^{-\alpha t}(\alpha t)^i}{i!},
\end{eqnarray}
which is what we would expect for the random graph with mean degree of $\alpha t$ in the large size limit~\cite{boll}. 
We can similarly write the link-space master equation for this model:
\begin{eqnarray}\label{eqn:lsacc}
\langle L_{i,j}(t+1)\rangle &=&L_{i,j}(t) +\frac{\langle m(t) \rangle \Theta_{i-1}(t) L_{i-1,j}}{X_{i-1}}\nonumber \\
{}&{}& +\frac{\langle m(t) \rangle \Theta_{j-1}(t) L_{i,j-1}}{X_{j-1}} -\frac{\langle m(t) \rangle \Theta_{i}(t) L_{i,j}}{X_{i}} \nonumber\\
{}&{}& -\frac{\langle m(t) \rangle \Theta_{j}(t) L_{i,j}}{X_{j}} + i \Theta_{j-1}P\{m(t)= i\}\nonumber \\
{}&{}&+j \Theta_{i-1}P\{m(t)= j\}.
\end{eqnarray}
The last two terms reflect the expected number of new $i\leftrightarrow j$ links being formed between the new node and existing nodes. Recalling that the normalized link-space matrix is given by $l_{i,j}(t) = L_{i,j}(t)/M(t)$ where the number of links, $M(t)$, at time $t$ can be approximated as $M(t)\approx \alpha t^2/2$ and making use of the previously derived, time-dependent degree distribution such that $P\{m(t)=j\}=c_j(t)$, Eq.~(\ref{eqn:lsacc}) can be rewritten as
\begin{eqnarray}\label{eqn:lsacc2}
\frac{\textrm{d}}{\textrm{d}t}\left(\frac{\alpha t^2l_{i,j}(t)}{2}\right)&=&\frac{(\alpha t)^2}{2}\big(l_{i-1,j}(t) +l_{i,j-1}(t) -2l_{i,j}(t) \big)\nonumber \\
{}&{}&+\frac{2i j c_i(t)c_j(t)}{\alpha t}.
\end{eqnarray}
The solution to Eq.~(\ref{eqn:lsacc2}) is found by means of ansatz (the choice of which is explained in Section~\ref{sec:cor}) and is
\begin{eqnarray}\label{ERLS}
l_{i,j}(t)&=&\frac{2e^{-2\alpha t}(\alpha t)^{i+j-2}}{(i-1)!(j-1)!}.
\end{eqnarray}
This is the normalized link-space matrix for a classical random graph with a mean degree of $\alpha t$ in the large size limit and this is illustrated in Fig~\ref{fig:ER} where a comparison to a simulation is also made. 

\section{Degree correlations and assortativity}
\label{sec:cor}
\begin{figure}
\begin{center}
\includegraphics[width=0.45\textwidth]{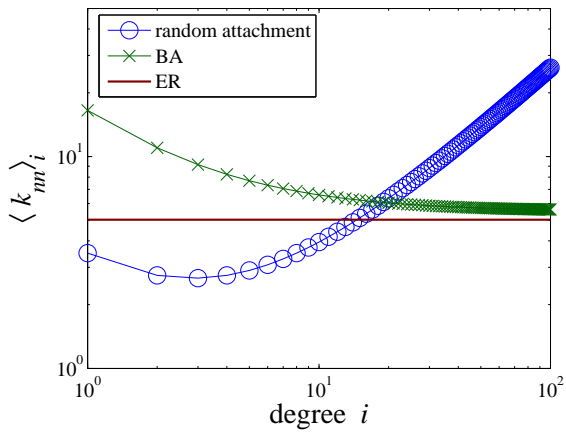}
\caption{ \label{fig:assort2} (Color Online) The mean degree of the nearest neighbours $\langle k_{nn}\rangle_i$ of nodes of degree $i$ as a function of $i$ for the analytic solutions to the random attachment algorithm (Section~\ref{subsec:RA}), the BA model (Section~\ref{subsec:BA}) and the ER random graph (Section~\ref{subsec:ER}).}
\end{center}
\end{figure}

The average nearest-neighbor degree $\langle{k}_{nn}\rangle_i$ of nodes of degree $i$ can be easily obtained from the link-space matrix and is given by 
\begin{equation}
\langle{k}_{nn}\rangle_i = \frac{\sum_{j=1}^{\infty} j L_{i,j}}{\sum_{j=1}^{\infty} L_{i,j}} =\frac{\sum_{j=1}^{\infty} j l_{i,j}}{\sum_{j=1}^{\infty} l_{i,j}}. 
\end{equation}
This is illustrated in Fig.~\ref{fig:assort2} for the steady-state solutions of the random attachment model, the BA model and the ER random graph. If the average nearest-neighbour degree $\langle{k}_{nn}\rangle_i$ is constant with respect to degree $i$, non-assortativity is implied. Certainly, the random attachment curve continues to increase, implying positive assortative mixing. The preferential attachment curve in Fig.~\ref{fig:assort2} appears to asymptote to a constant value (as was noted in \cite{equiluz:2002}) although the fact that it decreases initially illustrates that it is not perfectly non-assortative. This is a feature that would not be captured using a normalised Pearson's correlation coefficient which would suggest non-assortativity for this network~\cite{Newman:2003, David}.

\begin{figure*}
\begin{center}
\includegraphics[width=0.45\textwidth]{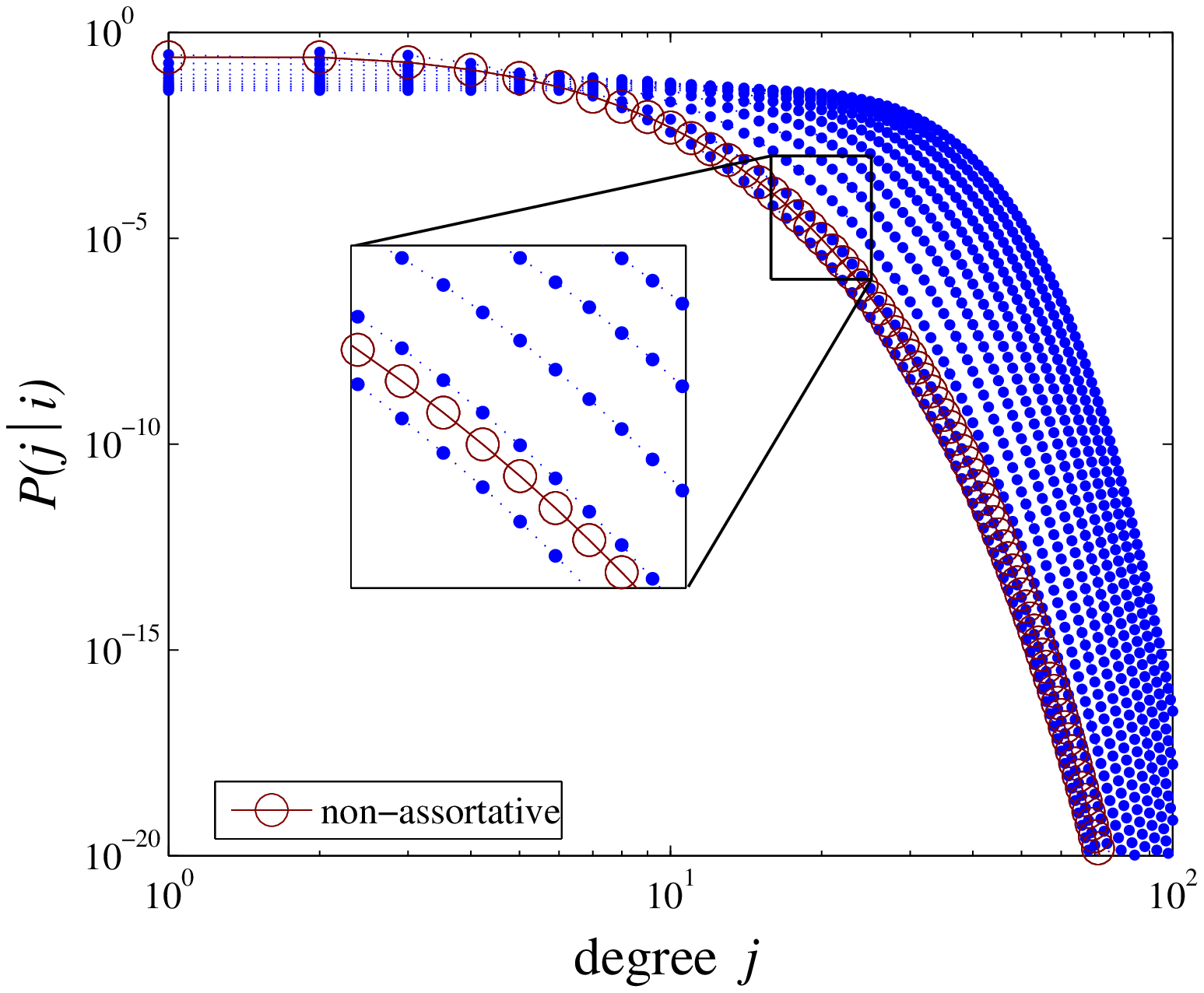}
\includegraphics[width=0.45\textwidth]{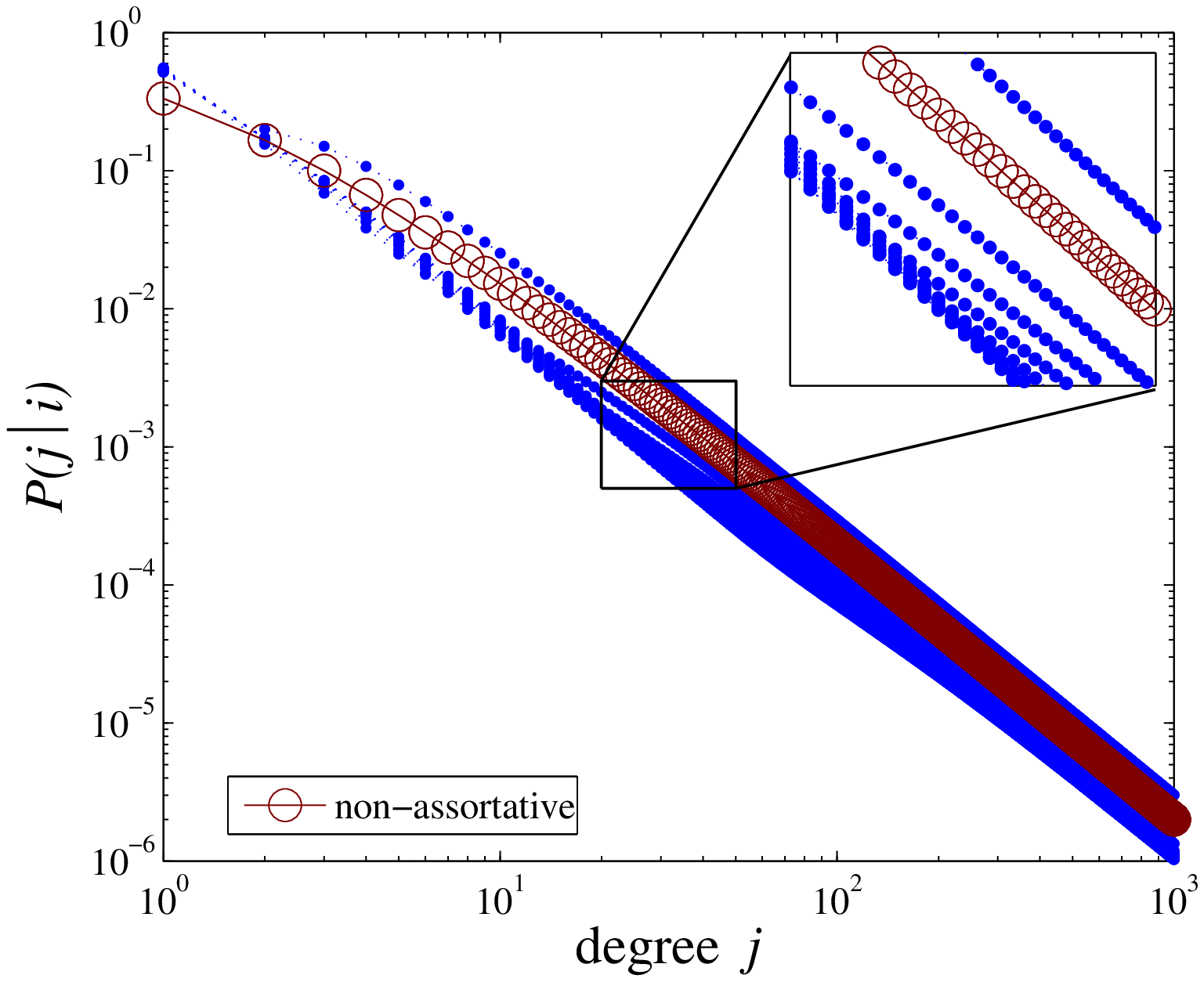}
\caption{ \label{fig:assort1} (Color Online) The conditional probability $P(j\mid i)$ that one end of a randomly selected link has degree $j$ given that the other end is of degree $i$ over the  range $i=\{1,6,11,16,...,51\}$. The left plot is for the random attachment algorithm and the right is preferential attachment. The criterion of non-assortativity of Eq.~(\ref{eqn:assort1}) is denoted with red circles.}
\end{center}
\end{figure*}

The link-space formalism allows us to address two-vertex correlations in a more powerful way in that we can calculate the conditional, two-point vertex degree distribution, $P(j|i)$ which has been traditionally difficult to measure~\cite{boguna:2003}. We can write this joint probability, i.e. the probability that a randomly chosen edge is connected to a node of degree $j$ given that the other end is connected to a node of degree $i$, in terms of the link-space matrix (normalized or not) as $P(j|i)= L_{i,j} / (i X_i) =  L_{i,j}/ \sum_{j=1}^{\infty} L_{i,j} =  l_{i,j}/ \sum_{j=1}^{\infty}  l_{i,j} $. For total number of edges aproximately equal to the total number of nodes, this can be approximated as $l_{i,j} / (i c_i)$. This is illustrated in Fig.~\ref{fig:assort1} for the scenarios of random and preferential attachment where the non-trivial nature of the correlations present within these networks is evident.

Consider selecting an edge at random in the network. If we then select one end of this edge at random, the probability that this node has degree $j$ will be proportional to $j$ since higher degree nodes have, by definition, more links connected to them than low degree nodes. Consider now only a subset of all edges with one end attached to a node of degree $i$. If there were no correlations present, the probability that the other end is attached to a node of degree $j$ is again proportional to $j$ ~\cite{Vazquez2}. The criterion of perfect non-assortativity for a network with equal numbers of nodes and links can be described as
\begin{eqnarray}\label{eqn:assort1}
P(j|i)&=& \frac{jc_j}{2}, \,\,\,\forall j.
\end{eqnarray}
This criterion is illustrated in Fig.~\ref{fig:assort1} for the random attachment and BA models and confirms that neither is perfectly non-assortative.

\subsection{Perfect non-assortativity}
\label{subsection:nonassort}
It is interesting to ask whether a perfectly non-assortative network can be generated. Assuming that such a network will not have equal numbers of nodes and links, we must rewrite the conditions of non-assortativity accordingly for total number of edges $M$: 
\begin{equation}
\label{eqn:assort3}
P(j|i)~=~ P(j)~=~\frac{j X_j}{2 M}~=~ \frac{j N c_j}{2 M}.
\end{equation}
Recalling our definition of the normalized link-space matrix that $l_{i,j} = L_{i,j}/M$, we can express this conditional probability $P(j|i)$ in terms of this matrix as
\begin{equation}\label{eqn:assort4}
P(j|i)~=~ \frac{L_{i,j}}{i X_i}~=~\frac{M l_{i,j}}{iNc_i}.
\end{equation}
As such, we can now write for $l_{i,j}$
\begin{eqnarray}
\label{eqn:assort5}
l_{i,j}&=& \bigg(\frac{N}{M}\bigg)^2 \frac{i c_i j c_j}{2}.
\end{eqnarray}
That is, for \emph{any} degree distribution, a normalized link-space matrix can be found which is representative of a perfectly non-assortative network \footnote{A network can be constructed for an arbitrary, normalized link-space matrix. A diagonal element $l_{i,i}$ can be simply obtained through the addition of fully connected components of $i+1$ nodes. Similarly an off-diagonal element $l_{i,j}$ can be generated through suitable rewiring of $i$-degree and $j$-degree cliques.}. This might provide an alternative to the network randomization technique of Maslov and Sneppen to provide a ``null model network" \cite{maslov:2002}.  The expression of Eq.~(\ref{eqn:assort5}) provides the ansatz solution for the normalized link-space matrix to the master equations for the ER random graph in Section~\ref{subsec:ER}.

\section{Decaying networks}\label{sec:decay}
Although somewhat counter-intuitive, it is possible to find steady states of networks whereby nodes and/or links are removed from the system. Aside from the obvious situation of having no nodes or edges left, we would like to investigate the possible existence of a network configuration whose link-space matrix and, subsequently, node degree distribution are static with respect to the decay process. The concept that decay processes are highly influential on a network's structure has been considered before although this has typically only been investigated in conjunction with simultaneous growth~\cite{May, Duplication, Laird}. Here we shall employ the link-space formalism to examine the effect of some simple, decay-only scenarios, specifically the two simplest cases -- random link removal and random node removal.

\subsection{Random Link Removal (RLR)}\label{subsec:rlr}
Consider an arbitrary network. At each timestep, we select a fixed number, $w$, of links at random and remove them. We shall implement the link-space formalism to investigate whether or not it is possible that such a mechanism can lead to stationary structure. Consider the link-space element $L_{i,j}(t)$ denoting the number of links from nodes of degree $i$ to nodes of degree $j$. Clearly this can be decreased if an $i \leftrightarrow j$ link is removed, i.e. a link that connects a degree $i$ node to a degree $j$ node. Also, if a  $k\leftrightarrow i$ link is removed and that $i$ node has further links to $j$ degree nodes, then those that were $i\leftrightarrow j$ links will now become $(i-1)\leftrightarrow j$, similarly for $k\leftrightarrow j$ links being removed. However, if the link removed is a $k\leftrightarrow (j+1)$ link and that $j+1$ degree node is connected to a degree $i$ node, then when the $j+1$ node becomes a degree $j$ node, the $(j+1)\leftrightarrow i$ link will become an $j\leftrightarrow i$ link, increasing $L_{i,j}$. Let us assume that we are removing links at random from the network comprising $N(t)$ vertices and $M (t)$ links. A non-random link selection process could be incorporated into the master equations using a probability kernel. The master equation for this process can be written in terms of the expected increasing and decreasing contributions:
\begin{eqnarray}\label{eqn:RLR1}
\langle L_{i,j}(t+1)\rangle &=& L_{i,j}(t) + w\sum_{k}\frac{L_{k,j+1}(t) j L_{i,j+1}(t)}{M(t) (j+1) X_{j+1}(t)}\nonumber\\
{}&{}&+w\sum_{k}\frac{L_{k,i+1}(t) i L_{i+1,j}(t)}{M(t) (i+1) X_{i+1}(t)}\nonumber\\
{}&{}&-w\sum_{k}\frac{L_{k,j}(t) (j-1) L_{i,j}(t)}{M(t) (j) X_{j}(t)}\nonumber\\
{}&{}&-w\sum_{k}\frac{L_{k,i}(t) (i-1) L_{i,j}(t)}{M(t) (i) X_{i}(t)}\nonumber\\
{}&{}&-\frac{wL_{i,j}(t)}{M(t)}.
\end{eqnarray}
This simplifies to
\begin{eqnarray}\label{eqn:RLR2}
\langle L_{i,j}(t+1)\rangle &=& L_{i,j}(t) + \frac{w j L_{i,j+1}(t)}{M(t)}\nonumber\\
{}&{}&+\frac{w i L_{i+1,j}(t)}{M(t)}-\frac{w (i-1) L_{i,j}(t)}{M(t)}\nonumber\\
{}&{}& -\frac{w (j-1) L_{i,j}(t)}{M(t)}-\frac{wL_{i,j}(t)}{M(t)}, 
\end{eqnarray}
where the last term refers to the physical removal of an $i\leftrightarrow j$ link.

To investigate the possibility of a steady state solution, we make a similar argument as before for growing networks but this time for the process of removing $w$ links per timestep:
\begin{eqnarray}
L_{i,j}(t)&=&l_{i,j}M(t)\nonumber\\
{}&=&l_{i,j}(M_0-wt),\nonumber\\
\frac{\textrm{d}L_{i,j}}{\textrm{d}t}&=&-wl_{i,j}\nonumber\\
{}&\approx & L_{i,j}(t+1)-L_{i,j}(t).
\end{eqnarray}
The expression of Eq.~(\ref{eqn:RLR2}) can be reduced to 
\begin{eqnarray}\label{eqn:lsrlr}
l_{i,j}&=&\frac{i l_{i+1,j}+j l_{i,j+1}}{i+j-2}.
\end{eqnarray}
As such, (twice) the fraction  of $1\leftrightarrow 1$ links, $l_{1,1}$,  does not reach a steady state. This might be expected, as the removal process for such links requires them to be physically removed as opposed to the process by which the degree of the node at one end of the link is being decreased. Therefore, the value $l_{1,1}$ increases in time. However, we can investigate the properties of the rest of the links in the network which do reach a steady state by neglecting these two node components (see Appendix~\ref{app:decay} for details). The normalized link-space matrix can be written for this system for $i+j\ne2$ as
\begin{eqnarray}\label{eqn:rlrfull}
l_{i,j}&=&\frac{A}{2^{i+j}}\frac{(i+j-3)!}{(i-1)!(j-1)!}.
\end{eqnarray}
The summation over this link-space matrix does not converge, i.e.  $\sum_i \sum_j l_{i,j} \rightarrow \infty$. However, we can still derive the degree distribution using Eq.~(\ref{eqn:dist}) and normalize this such that $\sum_i c_i =1$ (see Appendix~\ref{app:decay} for details):
\begin{eqnarray}\label{eqn:analrlr}
c_1&=&\frac{\log(2)}{1+\log(2)},\nonumber\\
c_i&=&\frac{1}{(1+\log(2))i(i-1)}\,\,\,\, i>1 .
\end{eqnarray}
Consequently, the first moment of the distribution diverges also, $\sum_i i c_i = \langle k \rangle \rightarrow \infty$.
It is interesting that the process of randomly removing links generates a network which exhibits power-law scaling of exponent $2$, corresponding to a degree distribution significantly over-skewed with respect to the preferential selection process of the BA growth algorithm (exponent 3).
By considering the average degree of the neighbours of nodes of degree $i$, as discussed in Section \ref{sec:cor}, we can see that the random link removal algorithm generates highly assortative networks.

\subsection{Random Node Removal (RNR)}
\label{subsec:RNR}
In a similar manner to Section \ref{subsec:rlr}, we will now discuss the possibility of creating such a steady-state structure via a process of removing nodes (with all of their links) from an existing network at the rate of $w$ nodes per timestep. Clearly, removing some node which has a link to an $(i+1)$-degree node which in turn has a link to a $j$-degree node can increase the number of $i\leftrightarrow j$ links in the system. The other processes which can increase or decrease the number of links from $i$ degree nodes to $j$ degree nodes can be similarly explained. We consider selecting a node of some degree $k$ for removal with some probability kernel $\Theta_k$ (as in the growing algorithms of Section \ref{sec:NSLS}). The master equation for such a process can thus be written for general node selection kernel:

\begin{eqnarray}\label{eqn:noderemovemaster}
\langle L_{i,j}(t+1)\rangle &=& L_{i,j}(t)~-~w\frac{\Theta_i(t) L_{i,j}(t)}{X_i(t)}-~w\frac{\Theta_j(t) L_{i,j}(t)}{X_j(t)}\nonumber\\
{}&{}&+~w\sum_k \frac{\Theta_k(t)}{X_k(t)} \bigg\{ L_{k,i+1}(t) \frac{i L_{i+1,j}(t)}{(i+1)X_{i+1}(t)}\nonumber\\
{}&{}&+~L_{k,j+1}(t) \frac{j L_{i,j+1}(t)}{(j+1)X_{j+1}(t)}\bigg\} \nonumber\\
{}&{}&-~w\sum_k \frac{\Theta_k(t)}{X_k(t)} \bigg\{ L_{k,i}(t) \frac{(i-1) L_{i,j}(t)}{iX_i(t)}\nonumber\\ 
{}&{}&~+ L_{k,j}(t) \frac{(j-1)L_{i,j}(t)}{j X_j(t)}\bigg\}.
\end{eqnarray}

We now specify a kernel for selecting the node to be removed, namely, the random kernel although the approach is general to any node selection procedure. Selecting a node purely at random leads to $\Theta_k=c_k$. The steady-state assumptions must be clarified slightly. As before
\begin{eqnarray}
L_{i,j}(t)&=&l_{i,j}M(t).
\end{eqnarray}
However, because we can remove more than one link (through removing a high degree node for example) we must approximate for $M(t)$. Using the random removal kernel, we can assume that on average, the selected node will have degree equal to the mean degree of the network, $\langle k(t)\rangle=\frac{2M(t)}{N(t)}$. We can use this to write the number of remaining links in the network as 
\begin{eqnarray}\label{eqn:rnrsteady}
L_{i,j}(t)&=&l_{i,j}(M(0)-w\langle k(t)\rangle) \nonumber\\
{}&=&l_{i,j}(M_0-\frac{2 w M (t)}{N(t)}t),\nonumber\\
\frac{dL_{i,j}}{dt}&=&-\frac{2wM(t)}{N(t)}l_{i,j}\nonumber\\
{}&\approx & L_{i,j}(t+1)-L_{i,j}(t).
\end{eqnarray}

Making use of the link-space identities and Eq.~(\ref{eqn:rnrsteady}),  the master equation Eq.~(\ref{eqn:noderemovemaster}) can be written in simple form for the steady state as
\begin{eqnarray}\label{eqn:rnr}
l_{i,j}&=&\frac{i l_{i+1,j}+ j l_{i,j+1}}{i+j-2}.
\end{eqnarray}
Clearly, this is  identical to Eq.~(\ref{eqn:lsrlr}) for the random link removal model of Section \ref{subsec:rlr} and, consequently, the analysis of the degree distribution will be the same too. It is interesting that the random node removal and random link removal lead to the same degree correlations.

\section{Mixture model}\label{sec:mm}
In this section we introduce a simple model that makes use of only local information about node degrees as microscopic mechanisms requiring global information are often unrealistic for many real-world networks~\cite{Vazquez}. It therefore provides insight into possible alternative microscopic mechanisms for a range of biological and social networks. The link-space formalism allows us to identify the transition point at which lower power-law exponent degree distributions switch to higher exponent distributions with respect to the BA preferential attachment model. While similar local algorithms have been proposed in the literature \cite{Krapivsky, Saramaki, Evans}, the strength of the approach followed here is the ability to describe the inherent degree-degree correlations. 

It is well known that a mixture of random and preferential attachment in a growth algorithm can produce power law-degree distributions with exponents $\gamma \in [3, \infty)$ when new links are only established between the new node and the existing network~\footnote{When new links are allowed between existing nodes, the exponent can be as low as $2$ for a mixture of random and preferential attachment~\cite{Evans}.}. It has often been assumed that a \emph{one step} random walk replicates linear preferential attachment \cite{Saramaki,Vazquez}. This is \emph{not} true. A one step random walk is in fact more biased towards high degree nodes than preferential attachment \footnote{A random walk of two steps in a network will be biased towards selecting low degree nodes. Previous studies have made use of this phenomenon when employing a two (or more) step random walk to select two (or more) nodes within the existing network for a new node to attach to with two (or more) links. The opposing biases effectively cancel such that the resulting algorithm can replicate preferential attachment. This was studied by Saram\"aki and Kaski \cite{Saramaki} and by Evans and Saram\"aki~\cite{Evans}. For a more detailed analysis see \cite{David}.} as can be easily seen by performing the procedure on a simple hub and spoke network. In this case, the probability of arriving at the hub tends to one for increasingly large networks rather than a half as would be appropriate for preferential attachment. We can use this bias to generate networks with degree distributions that have lower power-law exponent than the preferential attachment model, a phenomenon also observed by Evans and Saram\"aki~\cite{Evans}. A mixture of this approach with random attachment results in a simple model that can span a wide range of degree distributions. This one-parameter, growing network growth model, in which we simply attach a single node at each timestep with a single undirected link, does not require prior knowledge of the existing global network structure. 

\begin{figure*}
\begin{center}
$\begin{array}{@{} c  @{} | c @{} | c@{}}
a = 1 & a = 0.2 & a = 0\\
\epsfxsize=0.3\textwidth 
\epsffile{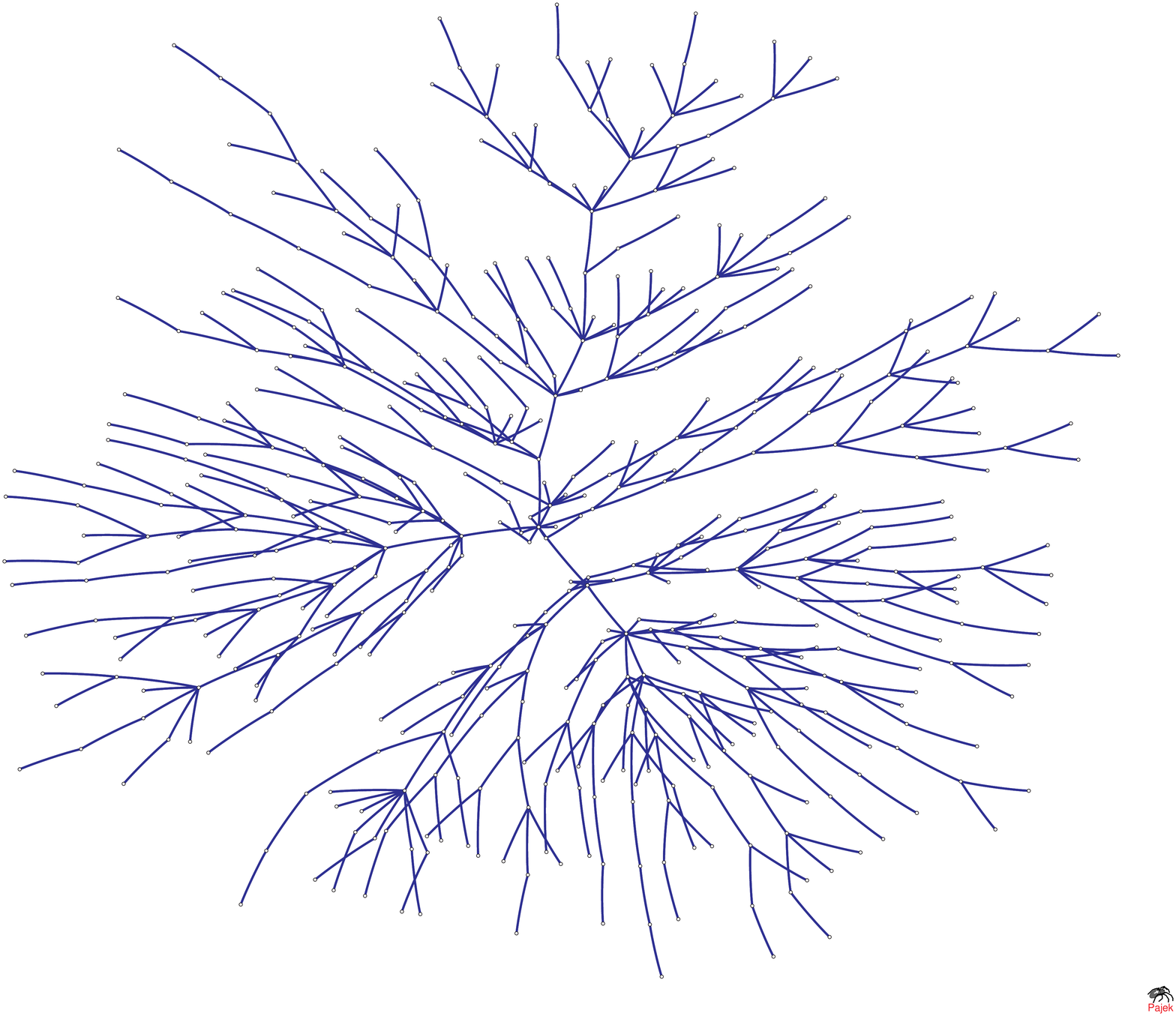} &
	\epsfxsize=0.3\textwidth
	\epsffile{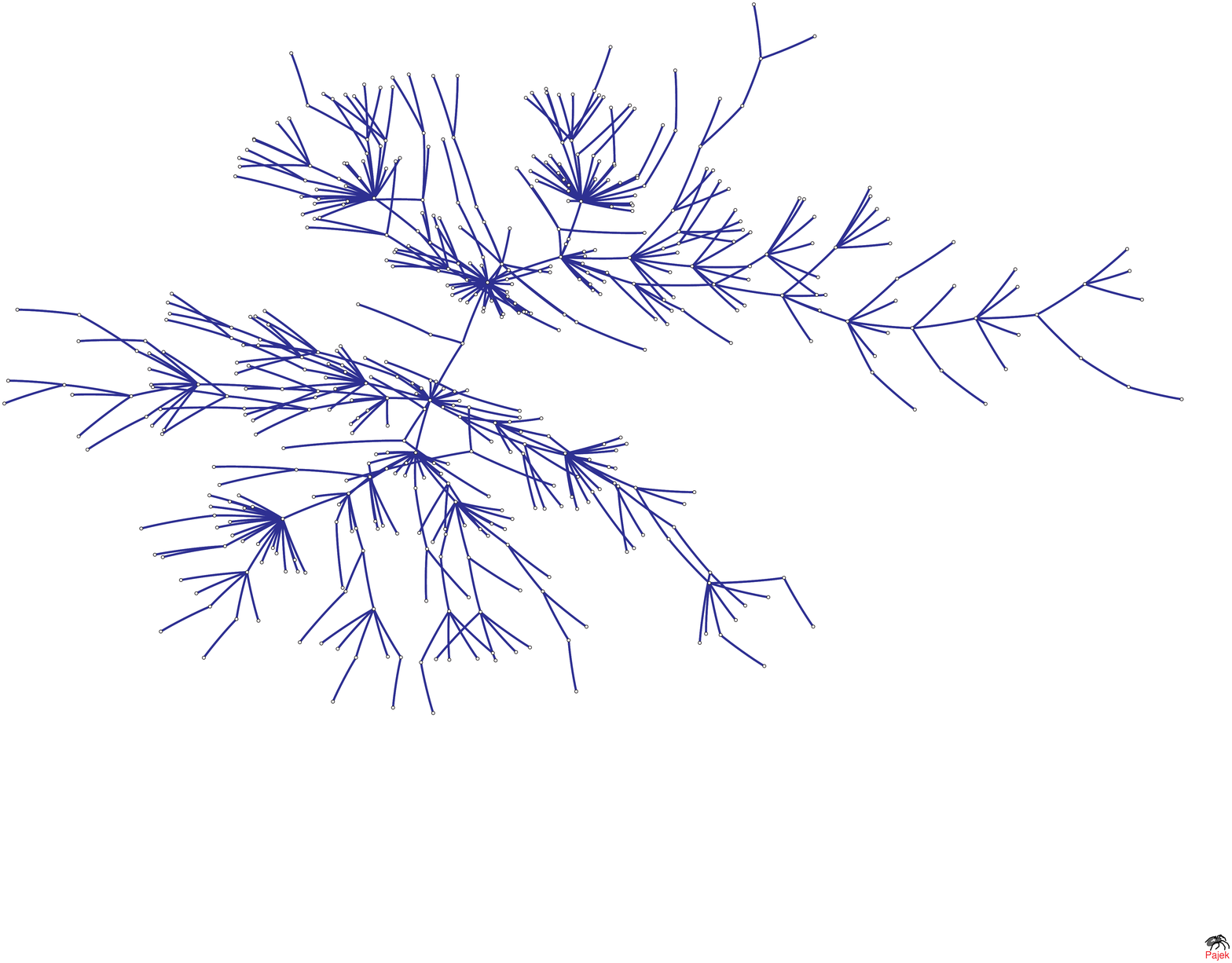} &
	\epsfxsize=0.3\textwidth
	\epsffile{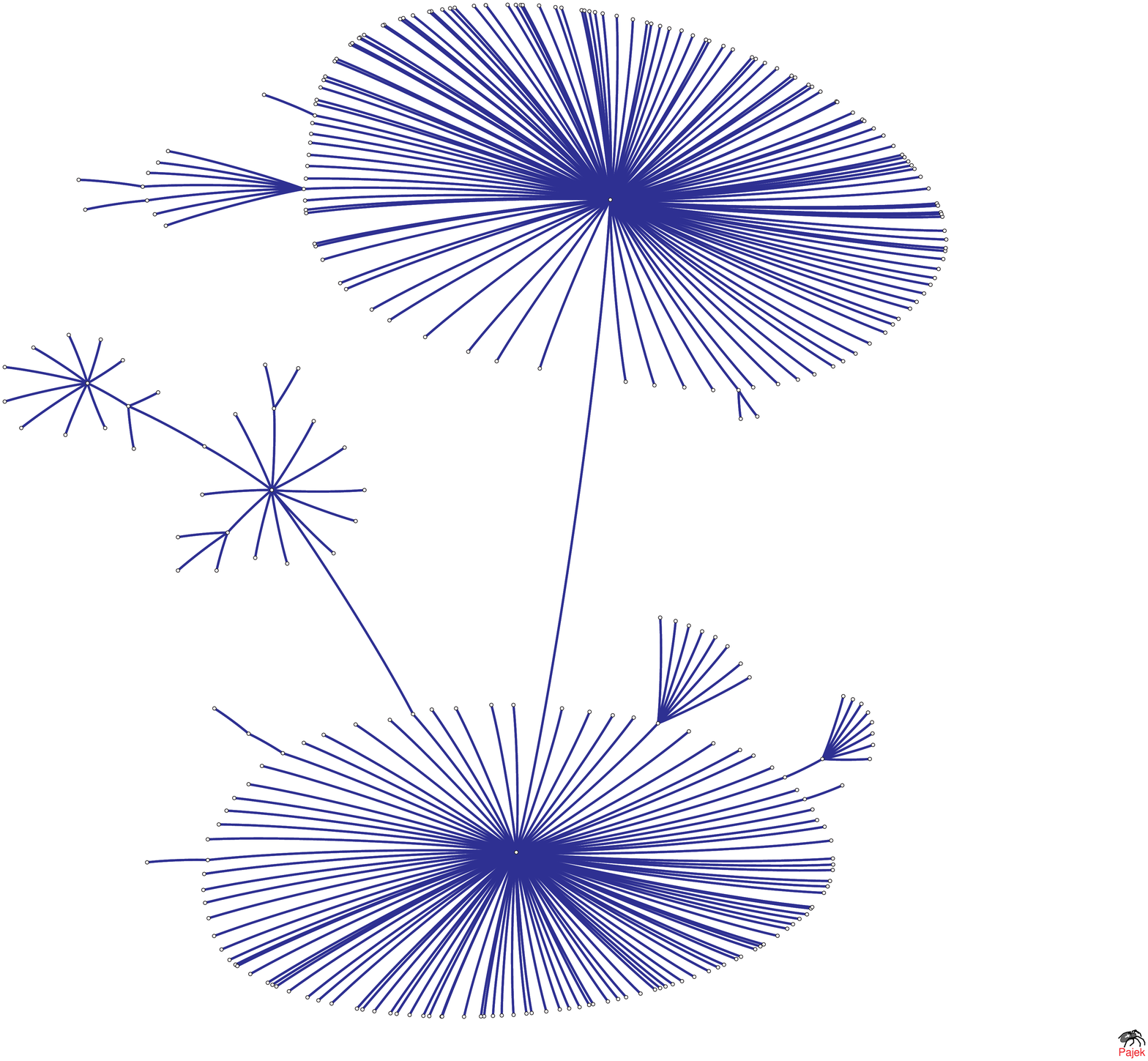}\\
	
	\epsfxsize=0.3\textwidth
	\epsffile{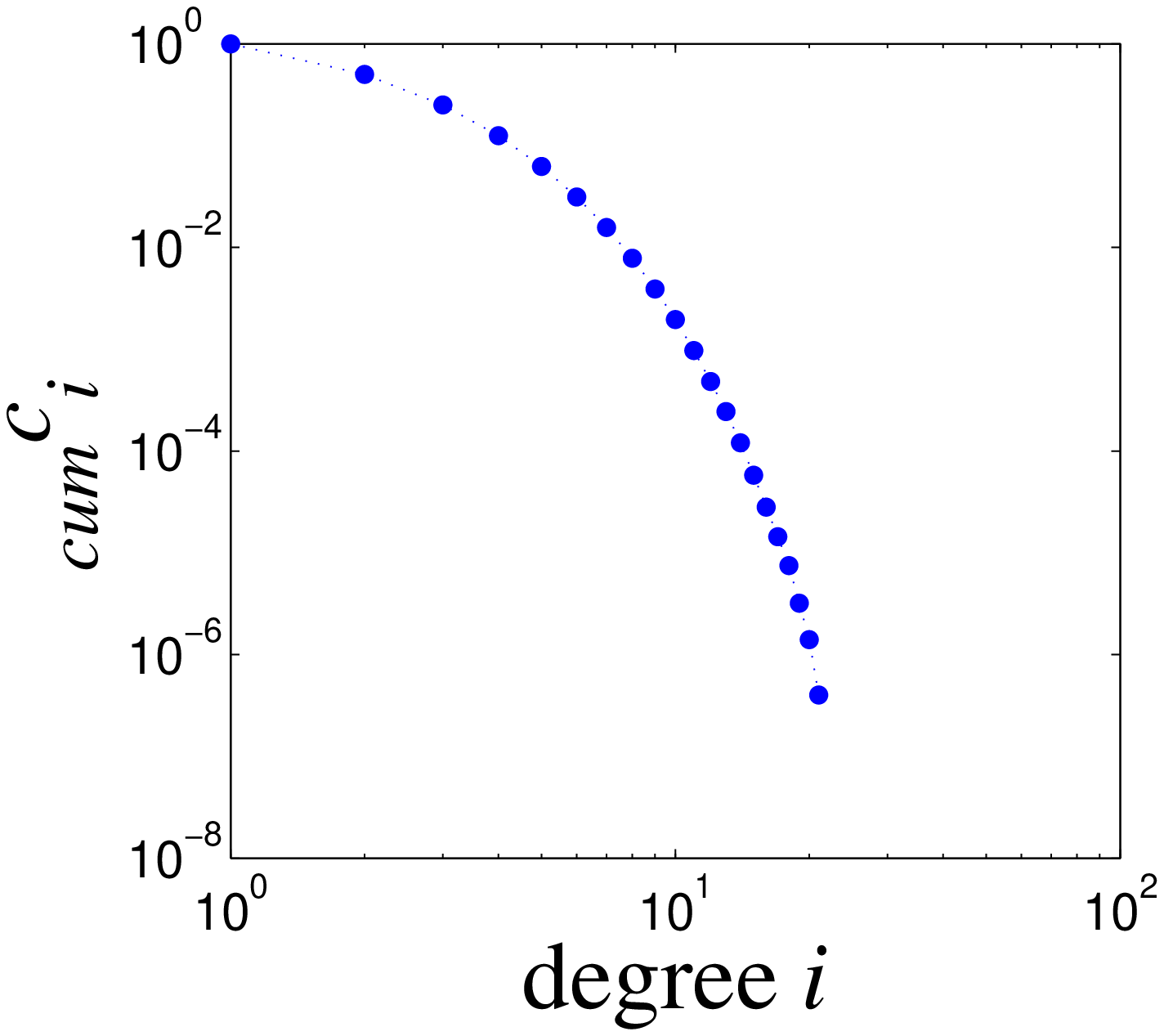} &	
	\epsfxsize=0.3\textwidth
	\epsffile{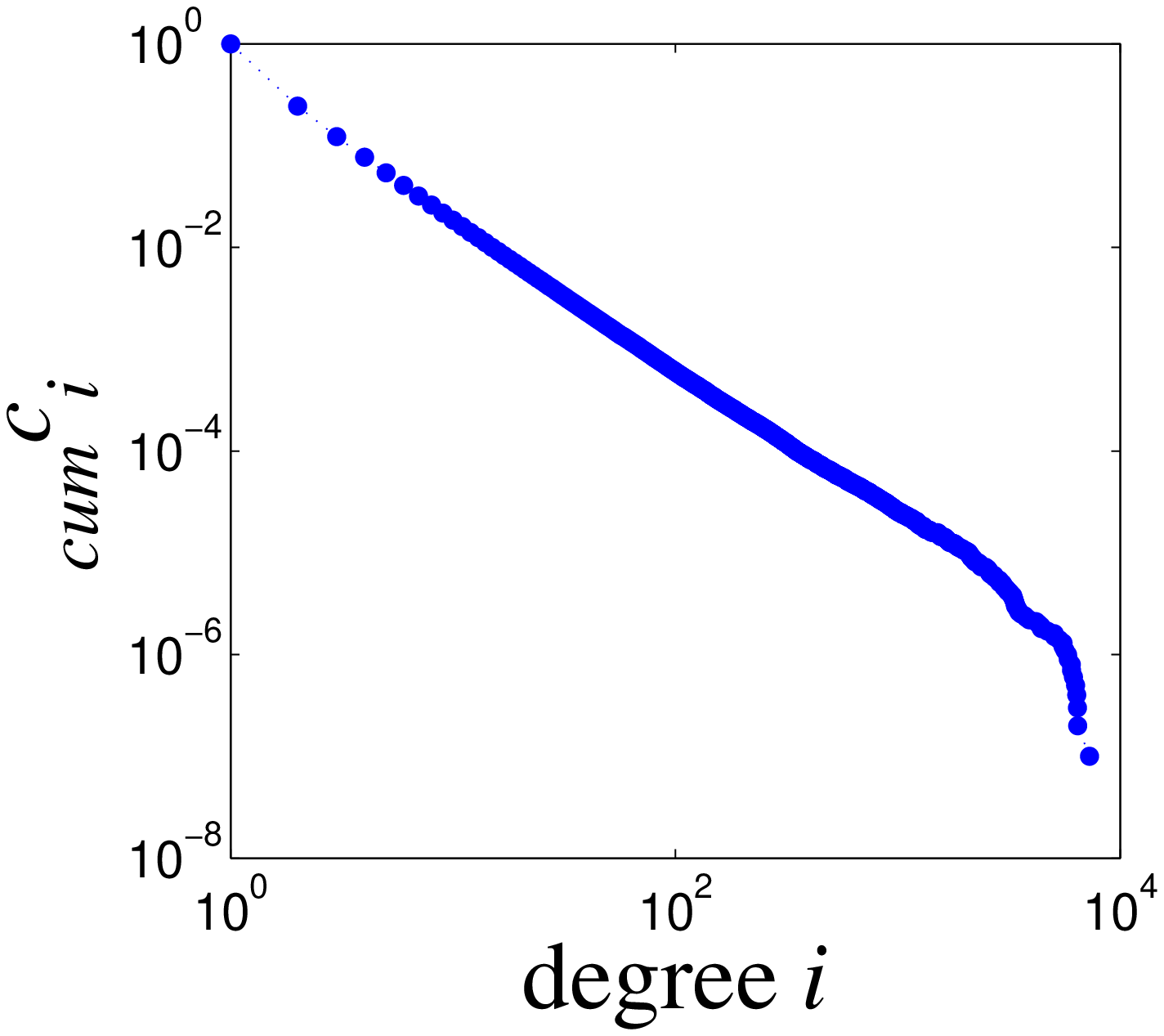} &
	\epsfxsize=0.3\textwidth
	\epsffile{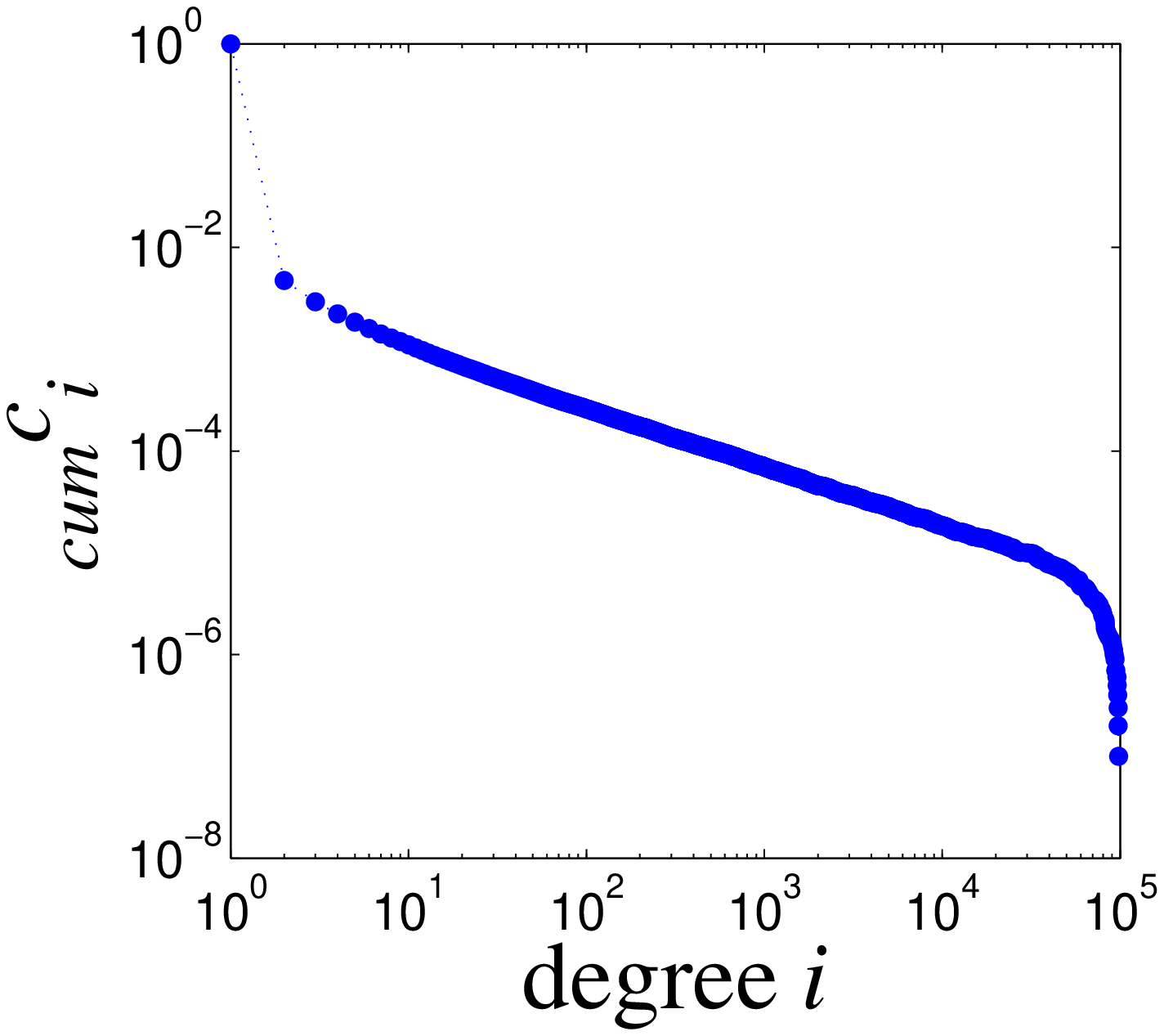} 
\end{array}$
\end{center}
\caption{\label{fig:exNN3plots} (Color Online) Top: Networks generated using the one-parameter, local information growth algorithm with $a =1$ (left),  $a=0.2$ (center), $a=0$ (right). Initial network seed comprised two nodes and one link. Networks drawn with Pajek~\cite{Batagelj98}. Bottom: Cumulative degree distributions for the same networks grown to $10^5$ nodes and ensemble-averaged over $100$ networks per $a$ value.}
\end{figure*}

\begin{figure}
\includegraphics[width=0.45\textwidth]{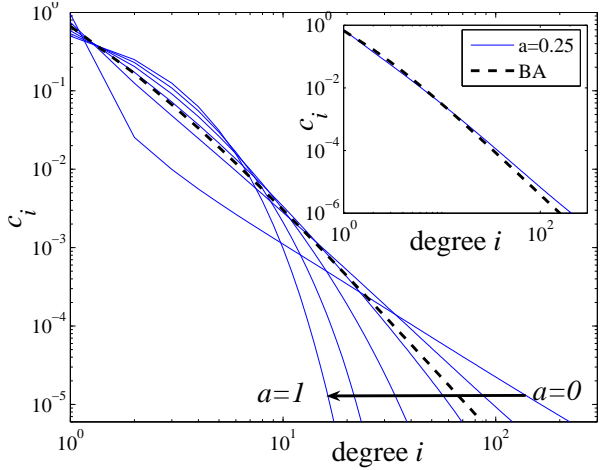}
\caption{ \label{fig:lsresults} (Color Online) Degree distributions 
generated using the link-space analysis of our one-step algorithm. Dashed line: results for the Barab\'asi-Albert (BA) \cite{Barabasi} preferential-attachment algorithm. Our one-step algorithm is closely resembles the BA results with parameter value $a=0.25$ (see inset).}
\end{figure}

The algorithm proceeds explicitly as follows: (i) pick a node $\kappa$ within the existing network at random; (ii) with probability $a$ make a link to that node; otherwise (iii) pick any of the neighbours of $\kappa$ at random and link to that node. Hence this algorithm resembles an object or `agent' making a short random-walk. This is very similar to the Growing Network with Redirection model introduced by Krapivsky and Redner~\cite{Krapivsky} except that the model here employs undirected links which necessitates a random (as opposed to deterministic) walk. Fig.~\ref{fig:exNN3plots} shows examples of the resulting networks and the corresponding (cumulative) degree distributions. Interestingly, $a=0$ yields a network that is dominated by hubs and spokes while $a = 1$ yields the random attachment network. Intermediate values of $a$ yield networks which are neither too ordered nor too disordered. For $a\sim 0.2$, the algorithm generates networks whose degree-distribution resembles the BA preferential-attachment network (see Figs.~\ref{fig:exNN3plots} and \ref{fig:lsresults}).

Our analysis of the algorithm starts by establishing the attachment kernel, $\Theta_i(t)$, which in turn requires properly resolving the one-step random walk. The link-space formalism provides us with an expression for the probability $P'_i(t)$ associated with performing a  random walk of length one and arriving at a degree $i$ node~\cite{David}:
\begin{equation}\label{eqn:neighbour}
P'_i(t)= \frac{X_1(t) L_{1,i}(t)}{N(t)  
X_1(t)}+\frac{X_2(t) L_{2,i}(t)}{2 N(t)   
X_2(t)}+\frac{X_3(t) L_{3,i}(t)}{3 N(t)  X_3(t)}+\ldots 
\end{equation}
This can also be written \cite{David} as $P'_i(t) =\frac{i X_i(t)}{N(t)}\langle\frac{1}{k_{nn}(t)}\rangle_i$
where the average is performed over the neighbours of nodes with degree 
$i$. Note that this quantity does {\em not} replicate 
preferential attachment, in contrast to what is commonly thought 
~\cite{Saramaki, Vazquez}.
Defining $\beta_i(t)$ as
\begin{equation}
\beta_i(t) \equiv \frac{1}{i~c_i(t)}\langle \frac{1}{k_{nn}(t)}\rangle_i = \frac{\sum_k 
\frac{L_{i,k}(t)}{k}}{\sum_k L_{i,k}(t)}
=\frac{\sum_k \frac{l_{i,k}(t)}{k}}{\sum_k l_{i,k}(t)} \ ,
\label{eqn:beta}
\end{equation}
yields 
\begin{equation}
\label{eqn:xnn}
\Theta_i(t)= a c_i(t)~+~(1-a)\beta_i(t) i c_i(t)  \ .
\end{equation}
To investigate a possible steady-state, we could substitute the above equations into Eqs.~(\ref{eqn:nsmaster}) and (\ref{eqn:linkspacemaster}). However the non-linear terms resulting from $\beta$ imply that a complete analytical solution for $l_{i,j}$ and $c_i$ is difficult. We leave this as a future challenge, but stress that our formalism  can be implemented in its non-stationary form Eq.~(\ref{eqn:lspace}) numerically by iteration with very good efficiency \cite{David}, yielding the degree distributions shown in Fig.~\ref{fig:lsresults}.

We can now use the link-space formalism to deduce the parameter value at which our algorithm yields the BA degree distributionin the long-time limit. At this value $a=a_c$, the node-degree distribution goes from lower to higher power-law exponents with respect to the BA model. For this parameter value, the attachment probability to nodes of various degrees is equal for both our mixture algorithm and the BA model. Using Eqs.~(\ref{eqn:SF}) and (\ref{eqn:xnn}), we have $\frac{i}{2}= a_c~+~(1-a_c)\beta_i i $, and hence for large $i$ this yields $a_c=1-\frac{1}{2\beta_i}$. We could then proceed to use the exact solution of the link-space equations for the preferential-attachment algorithm, in order to infer $\beta_i$ in the high $i$ limit. However, since $\beta_i$ can be expanded in terms of $l_{i,j}$ as shown in Eq.~(\ref{eqn:beta}) and $l_{i,j}$ decays very rapidly as $i,j$ become large, we can obtain a good approximation by using only the first two terms of Eq.~(\ref{eqn:sfexact}). This yields $\beta\approx0.66$. Hence, the critical value at which this simple model approximates the BA degree distribution is $a_c=0.25$, illustrated with the inset plot of Fig.~\ref{fig:lsresults}.

\section{Conclusion}\label{sec:conc}
We have developed a new formalism which accounts for degree-degree correlations in networks. We have employed the formalism to produce analytic solutions to the link-space matrix for the random attachment, BA and ER network models allowing detailed analysis of the degree correlations therein. We have introduced a cumulative implementation of the formalism which can be used to compare link-space matrices and might also be applied to empirical networks. We have shown that a perfectly non-assortative network can be generated with arbitrary degree distribution. We have also demonstrated the possiblility of a steady-state of the degree distribution and of the two-point degree correlations for simple decaying networks through deriving the static, normalized link-space matrix for such a process. Employing the link-space formalism allowed us to accurately describe a simple one-parameter network growth algorithm which is able to reproduce a wide variety of degree distributions without any global information about node degrees. We have used the framework to show that a one-step random walk does \emph{not} replicate preferential attachment except in the particular case of a perfectly non-assortative network. Indeed this in itself represents a criterion for perfect non-assortativity. Whilst the present paper focuses on introducing and illustrating the use of the link-space formalism for a variety of model networks, we stress that its applicability is far more general.

\textbf{Acknowledgements:} For interesting and helpful suggestions, the authors would like to thank T.S. Evans, M.A. Porter and one anonymous referee. D.M.D.S. is supported by the European Union (MMCOMNET), C.F.L. by a Glasstone Fellowship (Oxford), and J-P.O. by a Wolfson College Junior Research Fellowship (Oxford).

\appendix
\section{\label{app:RA}Exact solution of random attachment}
\begin{figure}
\begin{center}
\includegraphics[width=0.45\textwidth]{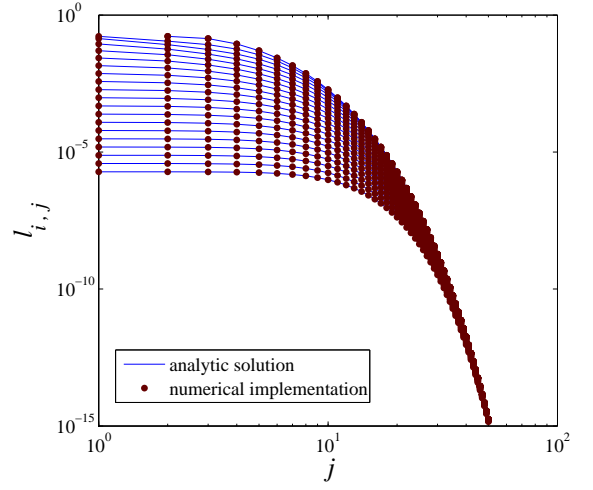}
\caption{ \label{fig:raanalvexact} (Color Online) Comparison of the numerically derived link-space matrix (markers) and the analytic solution (solid lines) for random node attachment for the first $20$ rows ($i$ values) of the link-space matrix. The numerical implementation populates the link-spcae matrix directly from the degree distribution and the steady-state link-space recurrence relations.}
\end{center} 
\end{figure}

For the random attachment algorithm, where one new node is added to the existing network with one new link at each timestep, the link-space master equation can be expressed for the steady state as
\begin{eqnarray}\label{eqn:app:ramaster2}
l_{i,j}&=& \frac{l_{i-1,j}+l_{i,j-1}}{3}, \ \ \ \ i,j~>1,\nonumber\\
l_{1,j}&=& \frac{c_{j-1}+l_{1,j-1}}{3},  \ \ \ \ j>1,\nonumber\\
l_{1,1}&=& 0.
\end{eqnarray}
It is easy to populate the link-space matrix numerically, just from the degree distribution $c_i = 2^{-i}$ obtained from solving the node-space recurrence relation and from the link-space master equation Eq.~(\ref{eqn:app:ramaster2}). At first glance, the solution to the master equation~\ref{eqn:app:ramaster2} would be of the form (check by substitution) 
\begin{eqnarray}
l_{i,j}&=& \frac{b}{4~2^{i+j}~3^{i+j}}.
\end{eqnarray}
However, the boundary conditions, which could be interpreted as influx of probability into the diffusive matrix, are such that this does not hold. We can actually solve the normalized link-space matrix for this model exactly. 
\begin{figure}
\begin{center}
\includegraphics[width=0.45\textwidth]{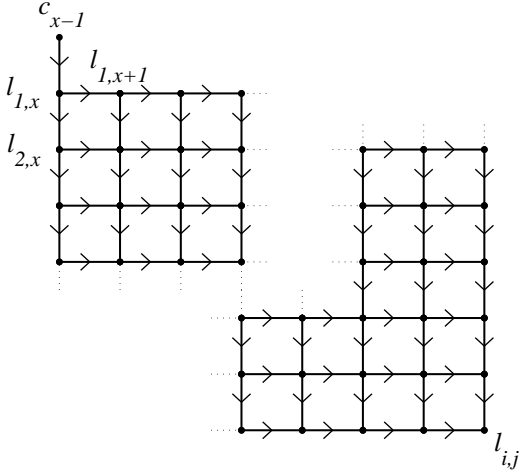}
\caption{ \label{fig:RAdiag} The paths of probability flux from $c_{x-1}$ influencing element $l_{i,j}$. Each arrow (step) represents a further factor of $\frac{1}{3}$.}
\end{center} 
\end{figure}
Consider the values $c_i$ as being influxes of probability into the top and left of the link-space matrix. We can compute the effect of one such element on the value in the matrix $l_{i,j}$. Each step in the path of probability flux reflects an extra factor of $\frac{1}{3}$. First, we will consider the influx effect from the top of the matrix as in Fig.~\ref{fig:RAdiag}. The total path length from influx $c_{x-1}$ to element $l_{i,j}$ is simply $i+j-x$. The number of possible paths between $c_{x-1}$ and element  $l_{i,j}$ can be expressed as ${j-x+i-1 \choose j-x}$. We can similarly write down the paths and lengths for influxes into the left hand side of the matrix. The first row (and column) can be described as 
\begin{eqnarray} 
l_{1,j}& =& \frac{c_{j-1}~+~l_{1,j-1}}{3}\nonumber\\
{}&=& \frac{2^{-(j-1)}~+~l_{1,j-1}}{3}\nonumber\\
{}&=& \sum_{k=1}^{j-1}\frac{1}{3^k 2^{j-k}}.
\end{eqnarray}
Subsequent rows can be similarly described. So, for  $i,j>1$ an element can be written:
\begin{eqnarray}
l_{i,j}&=& \frac{l_{i-1,j}+l_{i,j-1}}{3}\nonumber\\
{}&=&\sum_{x=2}^{j} \frac{  {i-1+j-x \choose j-x}} {3^{(i+j-x)}2^{(x-1)}} +
\sum_{x=2}^{i} \frac{{i-1+j-x\choose i-x }}{3^{(i+j-x)}2^{(x-1)}}. 
\end{eqnarray}
The results can be observed in Fig.~\ref{fig:raanalvexact} where a comparison to the numerically populated link-space matrix is made. 

\section{Exact solution of preferential attachment}
\label{app:BA}
In the BA model, when adding one new link with one undirected node per timestep, the steady-state solution of the node-space master equation leads to the degree distribution
\begin{eqnarray}
\label{eqn:app:NSpref}
c_{i}&=&\frac{4}{i(i+1)(i+2)}, 
\end{eqnarray} 
which can be checked easily by substitution. Substituting the attachment probability kernel $\Theta_i \approx i c_i / 2$ into the link-space master equation Eq.~(\ref{eqn:linkspacemaster}), we obtain the master equations for the link-space for this network growth algorithm:
\begin{eqnarray}\label{eqn:app:sfmaster3}
l_{i,j}&=&\frac{(i-1)l_{i-1,j}~+~(j-1)l_{i,j-1}}{2+i+j},\ \ \ \ i,j>1, \nonumber \\
l_{1,j}&=&\frac{(j-1)c_{j-1}~+~(j-1)l_{1,j-1}}{3+j}, \ \ \ \ j>1, \nonumber \\
l_{1,1}&=&0.
\end{eqnarray}
Again, it is easy to populate the matrix numerically just by implementing the node and link-space equations. 
At first glance, the solution to the master equation Eq.~(\ref{eqn:app:sfmaster3}) would be of the form (check by substitution) 
\begin{eqnarray}
l_{i,j}&=& \frac{A}{i(i+1)j(j+1)}, 
\end{eqnarray}
where $A$ is a constant. This would imply for the degree distribution 
\begin{eqnarray}\label{eqn:wrongdist}
c_i &=& \frac{A}{i^2(i+1)}. 
\end{eqnarray}
Summing over the entire node-space would give $A = \frac{6}{\pi^2-6}$. However, the boundary conditions (which could be interpreted as influx of probability into the diffusive matrix) are such that this solution does not hold. This is evident when comparing Eq.~(\ref{eqn:wrongdist}) with Eq.~(\ref{eqn:app:NSpref}), which compares to simulated networks well.

We can obtain an exact (although somewhat less pretty) solution by tracing fluxes of probability around the matrix and making use of the previously derived degree distribution. We can rewrite our link-space master equation, (\ref{eqn:app:sfmaster3}), in terms of vertical and horizontal components:
\begin{eqnarray}
l_{i,j}&=&\Psi_{i,j}l_{i-1,j} ~+~ \Upsilon_{i,j}l_{i,j-1}, \nonumber\\
l_{1,j}&=& \Upsilon_{i,j}\big(l_{1,j-1}~+~c_{j-1}\big),\nonumber\\
l_{1,1}&=&0, 
\end{eqnarray}
where 
\begin{eqnarray}
\Psi_{i,j} &=& \frac{\frac{\Theta_{i-1}}{c_{i-1}}}{1~+~\frac{\Theta_{i}}{c_{i}}~+~\frac{\Theta_{j}}{c_{j}}},\nonumber\\
\Upsilon_{i,j} &=& \frac{\frac{\Theta_{j-1}}{c_{j-1}}}{1~+~\frac{\Theta_{i}}{c_{i}}~+~\frac{\Theta_{j}}{c_{j}}}.
\end{eqnarray}

\begin{figure}
\begin{center}
\includegraphics[width=0.45\textwidth]{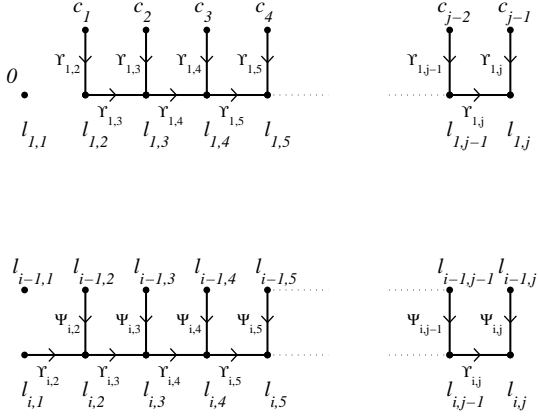}
\caption{ \label{fig:SFexact} The components of flux of probability around the link-space matrix. This is general to any attachment kernel. An element within the link-space matrix can be built up from contributing elements and the appropriate factors.}
\end{center} 
\end{figure}

By considering the probability fluxes as shown in Fig.~\ref{fig:SFexact}, we can write the individual elements in the link-space matrix as 
\begin{eqnarray}
l_{1,j}&=&\sum_{y=2}^j \bigg(c_{y-1} \prod_{x=y}^j \Upsilon_{1,x}\bigg), \nonumber\\
l_{i,j}&=&\sum_{y=2}^j \bigg(l_{i-1,y} \Psi_{i,y} \prod_{x=y+1}^j \Upsilon_{i,x}\bigg) + l_{i,1}\prod_{x=2}^j \Upsilon_{i,x} ,\nonumber\\
l_{1,1} &= &0.
\label{eqn:app:elements}
\end{eqnarray}

Note that we have yet to introduce the attachment probability kernels and the analysis so far is general.
Using the preferential attachment kernel, $\Theta_i \approx  \frac{i c_i}{2}$, we can write our component-wise factors for the master equation as
\begin{eqnarray}
\Psi_{i,j} &=& \frac{i-1}{i+j+2},\nonumber\\
\Upsilon_{i,j} &=& \frac{j-1}{i+j+2}.
\label{eqn:app:components}
\end{eqnarray}
Substituting Eq.~(\ref{eqn:app:components}) into Eq.~(\ref{eqn:app:elements}) and using the previously derived degree distribution of Eq.~(\ref{eqn:app:NSpref}) yields for the first row 
\begin{eqnarray}
l_{1,j}&=&\sum_{y=2}^j \bigg( \frac{4}{y(y-1)(y+1)} \prod_{x=y}^{j}\frac{x-1}{x+3}\bigg) \nonumber\\
{}&=&\frac{4(j-1)!}{(j+3)!}\sum_{y=2}^{j}(y+2) \nonumber\\
{}&=& \frac{2(j+6)(j-1)}{j(j+1)(j+2)(j+3)}.
\end{eqnarray}
Subsequent rows can be written as 
\begin{eqnarray}
l_{i,j} &=&l_{i,1}\frac{(j-1)!(3+i)!}{(2+i+j)!}\nonumber\\
{}&{}&~+~\sum_{y=2}^j l_{i-1,y}(i-1)\frac{(j-1)!}{(2+i+j)!}\frac{(1+i+y)!}{(y-1)!}\nonumber\\
{}&=& \frac{(j-1)!}{(2+i+j)!}\bigg\{(3+i)!l_{i,1},\nonumber\\
{}&{}&~+~(i-1)\sum_{y=2}^j l_{i-1,y} \frac{(1+i+y)!}{(y-1)!} 
\bigg\}.
\label{eqn:app:working1}
\end{eqnarray}
Rewriting this gives
\begin{eqnarray}
l_{i,j}&=& \frac{(j-1)!}{(2+i+j)!}\bigg\{K_i~+~E_{i-1,j}\bigg\}, 
\label{eqn:app:working2}
\end{eqnarray}
where the meaning of $K_{i}$ and $E_{i-1,j}$ follows from Eq.~(\ref{eqn:app:working1}). Clearly, we can write Eq.~(\ref{eqn:app:working2}) for $l_{i-1,y}$ as 
\begin{eqnarray}
l_{i-1,y}&=&\frac{(y-1)!}{(1+i+y)!}\bigg\{K_{i-1}~+~E_{i-1,y}\bigg\}. 
\label{eqn:app:working3}
\end{eqnarray}
Substituting Eq.~(\ref{eqn:app:working3}) into Eq.~(\ref{eqn:app:working1}) yields 
\begin{eqnarray}
l_{i,j}&=&  \frac{(j-1)!}{(2+i+j)!}\bigg\{K_i~+~(i-1)\sum_{y=2}^j\big(K_{i-1}~+~ E_{i-1,y}\big)\bigg\}. \nonumber\\
{}&{}&{}
\label{eqn:app:working4}
\end{eqnarray}
In order to solve this recurrence relation, we define an operator for repeated summation, $S^n_{j,y}$, such that 
\begin{eqnarray}
S_{j,y}(f(y)) &=& \sum_{y=2}^j f(y),   \nonumber\\
S^n_{j,y}(f(y)) &=& \sum_{y_n=2}^j \sum_{y_{n-1}=2}^{y_{n}} \sum_{y_{n-2}=2}^{y_{n-1}}\dots
\sum_{y_3=2}^{y_4} \sum_{y_{2}=2}^{y_{3}}\sum_{y=2}^{y_2}f(y). \nonumber \\
{}&{}&{}
\end{eqnarray}
The first subscript denotes the initial variable to be summed over and the second the final limit. 
A few examples of this operation will clarify its use:
\begin{eqnarray}
S^0_{j,y}(f(y))& =&f(y),\nonumber\\ 
S_{j,y}(1)&=& j-1  ,\nonumber\\
S^2_{j,y}(1) &=& S_{j,y}(y)-S_{j,y}(1) \nonumber\\
{} &=& \frac{1}{2}(j^2-j),\nonumber\\
S^3_{j,y}(1)&=&\frac{1}{6}(j^3-j),\nonumber\\
S_{j,y}(y)&=& \frac{j(j+1)}{2}-1 , \nonumber\\
S^2_{j,y}(y) &=& \frac{1}{6}(j^3+3j^2-4j).\nonumber\\
\end{eqnarray}
We can use this operator in our expression for the element $l_{i,j}$ in Eq.~(\ref{eqn:app:working4}) and expand to the easily derived value $E_{2,y}$:
\begin{eqnarray}
l_{i,j}&=& \frac{(j-1)!}{(2+i+j)!}\bigg\{ K_i~+~(i-1)S_{j,y}\big(K_{i-1}+E_{i-1,y}\big)\bigg\} \nonumber\\
{}&=& \frac{(j-1)!}{(2+i+j)!}\bigg\{ K_i+(i-1)K_{i-1}S_{j,y}(1) \nonumber\\
{}&{}&+ (i-1)(i-2)S^2_{j,y}\big(K_{i-2} +E_{i-2,y} \big) \bigg\}    \nonumber\\
{}&=& \frac{(j-1)!}{(2+i+j)!}\bigg\{ K_i+(i-1)K_{i-1}S_{j,y}(1) \nonumber\\
{}&{}&+ (i-1)(i-2)K_{i-2}S_{j,y}^2(1)\nonumber\\
{}&{}&+(i-1)(i-2)(i-3)K_{i-3}S_{j,y}^3(1)+...\nonumber\\
{}&{}&+(i-1)(i-2)(i-3)\times...\times2 K_2 S_{j,y}^{i-2}(1)\nonumber\\
{}&{}&  +(i-1)(i-2)\times....\times2S_{j,y}^{i-2}(E_{2,y}) \bigg\}. 
\end{eqnarray}
We can express $E_{2,y}$ in terms of the operator $S$ too:
\begin{eqnarray}
E_{2,y} = 4 S_{y,y}\big(2S_{y,y}(1)+S_{y,y}(y)  \big). 
\end{eqnarray}
The element $l_{i,j}$ can be expressed as 
\begin{eqnarray}
l_{i,j}&=& \frac{(j-1)!}{(2+i+j)!}\bigg\{ \sum_{m=2}^i \frac{(i-1)!}{(m-1)!}K_m S^{i-m}_{j,y}(1) \nonumber\\
{}&{}&+ 4 (i-1)!\big((2 S^i_{j,y}(1)+S^i_{j,y}(y)\big)\bigg\}. \nonumber
\label{eqn:app:working6}
\end{eqnarray}
This form is somewhat obtuse as the calculation of the operator values is less than obvious.
However, we can transform to a more easily interpreted operator $W(n)$ analogous to $S$ but with different limits such that  
\begin{eqnarray}
W_{j,y}(f(y)) &=& \sum_{y=1}^j f(y),\nonumber\\
W^n_{j,y}(f(y)) &=& \sum_{y_n=1}^j \sum_{y_{n-1}=1}^{y_{n}} \sum_{y_{n-2}=1}^{y_{n-1}}\dots
\sum_{y_3=1}^{y_4} \sum_{y_{2}=1}^{y_{3}}\sum_{y=1}^{y_2}f(y). \nonumber \\
{}&{}&{}
\end{eqnarray}
Whilst, at first glance, it looks like little progress has been made, we only need evaluate the repeated operation on initial function $f(y)=1$.  This is exactly solvable and we can rewrite in terms of a function $G(n)$, dropping the superfluous $y$ subscripts:
\begin{eqnarray}
\label{eqn:app:inductive}
G_j(n)&=& W^n_{j}(1), \nonumber\\
G_j(n)&=& \left\{\begin{array}{l r}
\frac{(j+n-1)!}{n!(j-1)!}&\textrm{for $n\ge0$}\\
0 & \textrm{for  $n<0.$} \end{array}\right.
\end{eqnarray}
The inductive proof associated with Eq.~(\ref{eqn:app:inductive}) can be understood by path counting for some repeating binomial process and is an intrinsic property of the combinatorial choose coefficient. Consider a repeated coin toss over $x$ steps. The number of ways of achieving $n+1$ successes after these $x$ iterations is ${x \choose n+1}$. Now, the occurrence of this last success could have happened on the $(n+1)^{\textrm{th}}$ iteration or the following one, or any of the subsequent iterations till the $x^{\textrm{th}}$ one. For the last successful outcome to occur on the $m^{\textrm{th}}$ step, $n$ successful outcomes must have occurred in the previous steps. The number of ways this could have occurred is ${m-1 \choose n}$. Clearly, summing over all possible $m$ values, the total possible paths resulting in $n+1$ successes must equate to ${x \choose n+1}$. 

\begin{figure}
\begin{center}
\includegraphics[width=0.45\textwidth]{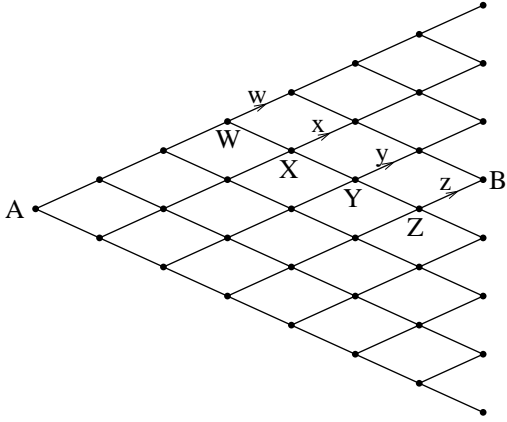}
\caption{ \label{fig:choosediag} A binomial process over seven steps. The number of paths between $A$ and $B$ can be expressed as the sum of the paths $A\rightarrow W$, $A\rightarrow X$, $ A\rightarrow Y$ and $A\rightarrow Z$.}
\end{center} 
\end{figure}

For clarity, this is depicted in Fig.~\ref{fig:choosediag}. To  reach point $B$ from $A$ in the binomial process, one of the steps $w,x,y$ or $z$ must be traversed, after which there is only one route to $B$. Consequently, the number of paths between $A$ and $B$ utilising step $w$ is the same as the number of paths between $A$ and $W$. Similarly, the number of paths between $A$ and $B$ utilising step $x$ is the same as the number of paths between $A$ and $X$ and so on. The number of paths between $A$ and $B$ can be built expressed as the sum of the paths $A\rightarrow W$, $A\rightarrow X$, $A\rightarrow Y$ and $A\rightarrow Z$. 

\begin{figure}
\begin{center}
\includegraphics[width=0.45\textwidth]{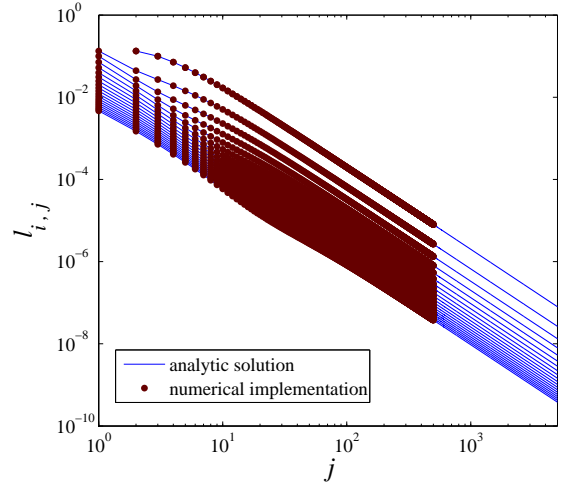}
\caption{ \label{fig:sfanalvexact} (Color Online) Comparison of the numerically derived link-space matrix (markers) and the analytic solution (solid lines) for preferential attachment. The numerical implementation populates the matrix from the solution of the degree distribution and the link-space recurrence equations.}
\end{center} 
\end{figure}

We incorporate this behaviour into our proof for the solution of $G_j(n)$.
\begin{eqnarray} 
G_j(n) &=& {j+n-1 \choose n}, \nonumber\\
G_j(n+1)&=&\sum_1^j G_j(n) \nonumber\\
{}&=&\sum_1^j~{j+n-1 \choose n} \nonumber\\
{}&=&{j+n \choose n+1}.
\end{eqnarray}
As $G(1)={j \choose 1}=j$, this inductive proof holds for all $n$ and the following relations hold:
\begin{eqnarray}
S^n(1)&=& G(n)-G(n-1),\nonumber\\
S^n(j)&=& G(n+1)-G(n-1).
\end{eqnarray}
We can now write the element in our link-space matrix for preferential attachment exactly as our function $G(n)$ is easily evaluated as
\begin{eqnarray}
l_{i,j}& =& \frac{4(j-1)!(i-1)!}{(j+i+2)!}\Bigg(G(i+1)+2G(i)-3G(i-1) \nonumber\\ {}&{}&+~\frac{1}{2}\sum_{k=2}^{i}(k-1)(k+6)\Big(G(i-k)-G(i-k-1)\Big)\Bigg). \nonumber\\
\end{eqnarray}
A comparison of this solution is made to a numerically-populated link-space matrix in Fig.~\ref{fig:sfanalvexact}.
\section{Degree distribution of decaying network}
\label{app:decay}
The link-space master equation for the steady state of the decaying networks of Section \ref{sec:decay} is written as 
\begin{eqnarray}\label{eqn:app:lsrlr}
l_{i,j}&=&\frac{i l_{i+1,j}+j l_{i,j+1}}{i+j-2}. 
\end{eqnarray}
The number of $1\leftrightarrow 1$ links does not reach a steady state for this process. However, we can investigate the properties of the rest of the links in the network which do reach a steady state by neglecting these two-node components. In a similar manner to~\cite{Krapivsky} we can make use of a substitution to find a solution to this recurrence equation, namely, for $i+j\ne2$ as 
\begin{eqnarray}\label{eqn:landm}
l_{i,j}&=& r_{i,j}\frac{(i+j-3)!}{(i-1)!(j-1)!}. 
\end{eqnarray}
From this, we can obtain
\begin{eqnarray}
r_{i,j}&=&r_{i+1,j}+r_{i,j+1}. 
\end{eqnarray}
This has the simple solution:
\begin{eqnarray}\label{eqn:msolution} 
r_{i,j}&=&\frac{A}{2^{i+j}}.
\end{eqnarray}
The link-space for this system can be written for $i+j\ne2$ as 
\begin{eqnarray}\label{eqn:app:rlrfull}
l_{i,j}&=&\frac{A}{2^{i+j}}\frac{(i+j-3)!}{(i-1)!(j-1)!}. 
\end{eqnarray}
Although the summation over this link-space matrix diverges, $\sum_i \sum_j l_{i,j} \rightarrow \infty$, we can use the form of Eq.~(\ref{eqn:app:rlrfull}) to infer the shape of the degree distribution. Recalling that
\begin{eqnarray}\label{eqn:ci3}
c_i&=& \frac{M}{N}\frac{\sum_k l_{i,k}}{i}, 
\end{eqnarray}
we can write for the degree distribution, neglecting the two node components in the network, as 
\begin{eqnarray}\label{eqn:rlrdegdist1}
c_1&=&\frac{M A}{N}\sum_{k=2}^{\infty}\frac{(k-2)!}{(k-1)! 2^{1+k}}, \nonumber \\
c_i&=&\frac{M A}{Ni}\sum_{k=1}^{\infty}\frac{(i+k-3)!}{(i-1)!(k-1)! 2^{i+k}}, \ \ \ \ i>1. 
\end{eqnarray}
The $c_1$ value can be solved by considering the Maclaurin Series of the function $\log(1+x)$ (as a historical note, this was done by Mercator as early as 1668) and evaluating for $x=-\frac{1}{2}$:
\begin{eqnarray}
\log(1+x)&=&\sum_{k'=1}^{\infty}\frac{(-1)^{k'}}{k'} x^{k'},\nonumber\\
\log\bigg(\frac{1}{2}\bigg)&=& -\sum_{k'=1}^{\infty}\frac{1}{k'~2^{k'}}\nonumber\\
{}&=&-\log(2).
\end{eqnarray}
Rearranging Eq.~(\ref{eqn:rlrdegdist1}) and letting $k=k'+1$ yields 
\begin{eqnarray}
c_1&=&\frac{M A}{N}\sum_{k=2}^{\infty}\frac{1}{(k-1) 2^{1+k}}\nonumber\\
{}&=&\frac{M A}{N}\sum_{k'=1}^{\infty}\frac{1}{k'2^{k'+2}}\nonumber\\
{}&=&\frac{M A}{4N}\log(2). \nonumber \\
\end{eqnarray}
For node degree greater than one, the solution of (\ref{eqn:rlrdegdist1}) requires a simple proof by induction. Consider the function $Q(i')$ defined for positive integer $i'$ as 
\begin{eqnarray}
Q(i')&=& \sum_{k'=0}^{\infty} \frac{{i'+k' \choose k'}}{2^{k'+i'}}.
\end{eqnarray}
We can write $Q(i'+1)$ with similar ease as 
\begin{eqnarray}
Q(i'+1)&=& \sum_{k'=0}^{\infty} \frac{ {i'+k'+1 \choose k'}}{2^{k'+i'+1}}. 
\end{eqnarray}
This leads to 
\begin{eqnarray}
2Q(i'+1)-Q(i')&=& \sum_{k'=0}^{\infty}\bigg({i'+k'+1 \choose k'}- {i'+k'\choose k'} \bigg)\frac{1}{2^{k'+i'}}\nonumber\\
{}&=& \sum_{k'=1}^{\infty} {i'+k' \choose k'-1} \frac{1}{2^{k'+i'}}. 
\end{eqnarray}
A quick substitution of $k''=k'-1$ leads to:
\begin{eqnarray}
2Q(i'+1)-Q(i')&=&\sum_{k''=0}^{\infty} {i'+k''+1 \choose k''} \frac{1}{2^{k''+i'+1}}\nonumber\\
{}&=&Q(i'+1),\nonumber\\
\Rightarrow~~Q(i'+1)&=&Q(i'),\nonumber\\
Q(1)&=&2\nonumber\\
{}&=&Q(i'), 
\end{eqnarray}
which holds of for all $i'$. Rearranging Eq. ~(\ref{eqn:rlrdegdist1}) and using some simple substitutions, $i'=i-2$ and $k'=k-1$, we can derive the following:
\begin{eqnarray}
c_i&=&\frac{M A}{Ni}\sum_{k=1}^{\infty}\frac{(i+k-3)!}{(i-1)!(k-1)! 2^{i+k}}\nonumber\\
{}&=&\frac{M A}{Ni(i-1)}\sum_{k=1}^{\infty}\frac{(i+k-3)!}{(i-2)!(k-1)! 2^{i+k}}\nonumber\\
{}&=&\frac{M A}{Ni(i-1)}\sum_{k'=0}^{\infty}\frac{ {i'+k' \choose k'}}{ 2^{i'+k'+3}}\nonumber\\
{}&=&\frac{M A}{4N}\frac{1}{i(i-1)}. 
\end{eqnarray}
We can normalize this degree distribution such that $\sum_i c_i = 1$ and so for the decaying network without the two node components (the $1\leftrightarrow 1$ links), the degree distribution can be expressed as 
\begin{eqnarray}\label{eqn:app:analrlr}
A&=& \frac{4N}{M (1+\log(2))},\nonumber\\
c_1&=&\frac{\log(2)}{1+\log(2)},\nonumber\\
c_i&=&\frac{1}{(1+\log(2))i(i-1)}, \ \ \ \ i>1.
\end{eqnarray}

\end{document}